\documentstyle[12pt]{article}
\begin{document}
\author{{\normalsize\bf Yu.A.Markov \thanks{e-mail:markov@icc.ru}
and M.A.Markova}}
\title{The problem of nonlinear Landau damping in quark-gluon
plasma}
\date{\it Institute of System Dynamics\\
and Control Theory Siberian Branch\\
of Academy of Scienses of Russia,\\
P.O. Box 1233, 664033 Irkutsk, Russia}
\thispagestyle{empty}
\maketitle

\[
{\bf Abstract}
\]

On the basis of the semiclassical kinetic equations for
quark-gluon plasma (QGP) and Yang-Mills equation, the
generalized kinetic equation for waves with regard to its
interaction is obtained. The physical mechanisms defining
nonlinear scattering of a plasmon by QGP particles are
analyzed. The problem on a connection of nonlinear Landau
damping rate of longitudinal oscillation with damping rate,
obtained on the basis of hard thermal loops approximation, is
considered.

\vspace{9cm}

{\bf 1. INTRODUCTION}

In recent 15-20 years, a theoretical
investigations of properties of quark-gluon plasma has been of great interest.
It is connected with intensive looking for
a QGP in the experiments with collision of ultrarelativistic heavy ions.

Two methods to study of the nonequilibrium phenomena in a
QGP are used: method of temperature Green functions
and kinetic approach. Significant progress has been achieved in
the development of the first method. The effective perturbative theory
was constructed in the papers by Pisarski and Braaten, Frenkel and Taylor [1]
on the basis of the resummation of
so-called hard-thermal loops (HTL's),
and the problem on the sign and gauge dependence of the
damping rate of the long wavelength excitations of QGP was solved.
The progress of the thermal QCD makes possible a new
view at the existing of kinetic theory of QGP, the basis of which was laid
in the papers [2] by Heinz, Elze, Vasak and Gyullasy, Mrowczynski and others,
and it has given impetus to its further development.

In spite of the fact that the language and methods appear very
different, there are close similarities
between HTL approach and transport theory. Originally the
kinetic theory was used by Silin to derive HTL in the photon's
self-energy [3]. HTL's in the quark and gluon self-energies
can be computed similarly. Moreover, Kelly, Liu, Lucchesi
and Manuel [4] have shown that the generating functional of HTL's (with
an arbitrary number of soft external bosonic legs) can be derived
from the classical kinetic theory of QGP. This points clearly to the classical
nature of the hard thermal effects.

A further step in development of kinetic theory was made
by Blaizot and Iansu [5]. In contrast to the early papers on transport
theory of QGP [2], these authors use from outset
the ideas developed in thermal QCD in deriving of the kinetic equations.
The equations obtained by
them isolate consistently the dominant terms in
the coupling constant $g$ in
hierarchy of equations which describe the response of plasma to weak and
slowly varying disturbances, and encompass all HTL's. However, here
it should be noted that if the influence of the average fermionic field
is neglected, then the expression for current induced by
soft gauge fields, obtained in [6]
(and the nonlinear equation of motion, connected with it)
fully
coincide with the corresponding
expression obtained in [4] from usual classical kinetic theory on the
basis of consistent expansion of distribution function in powers of the
coupling constant. In spite of
the fact that intermediate approximation schemes in which
these equations were derived, mix leading and nonleading
contributions with respect to the powers of $g$ and
so, are not entirely consistent, this, somewhat justifies use of the
classical and semiclassical kinetic equation found in [2].

Such close interlacing of two methods of investigation of nonequilibrium
phenomena in QGP leads to the question: can one
calculate whether the damping rate of bosonic modes corresponding to the hard thermal
result [7], remaining whithin the framework of classical (semiclassical)
kinetic theory only? Xiaofei and Jiarong [8] were the first
to put this question. Because of obtained results, they
have given a positive answer.

As was shown by Heinz and Siemens [9],
in linear approximation the Landau damping is absent in QGP.
In fact, the only mechanism, with which one can associate
the damping following from the kinetic theory with damping from
HTL approach, is so-called nonlinear Landau damping. It
bound up with the nonlinear effects of interaction of waves and particles
in QGP. The multiple time-scale method which has proved successful in studying
the nonlinear properties of electromagnetic plasma (EMP) [10],
was used in [8] for determination of this association. By means of this
method the nonlinear shift of an eigenfrequency of longitudinal oscillations
in the temporal gauge, the imaginary part of which defines required nonlinear
Landau damping rate, had been obtained by Xiaofei and Jiarong. Futher, the limiting
expression of the derived damping rate for ${\bf k}=0$-mode was obtained, and
numerical computations for approximate estimate were performed. The value derived
by this means is in close agreement with similar numerical one obtained
by Braaten and Pisarski [7] on the basis of effective perturbative theory.

However, under close examination of above-mentioned paper we found certain
inaccuracies in computations, which were of both principle and
nonprinciple character.
The elimination of these inaccuracies finally leads not only to a numerical
modification of the limiting value of nonlinear
Landau damping rate obtained in [8],
but what is more important, it changes the sign of obtained expression
(this subject will be considered in detail in Sec.10 below). This points
to some prematurity of the statements in [8] on obtained connection
between nonlinear Landau damping rate and damping rate from the
HTL-approach.

Moreover, the physical mechanisms lying in the basis of the
process of nonlinear scattering of longitudinal waves by QGP particles
has not been revealed by Xiafei and Jiarong.
The contribution to this process from the effect of
self-action of the gauge field was dropped, although in a "soft" region of
excitations it is of the same order as the contributions, bound up with
the effects of the medium. An important problem of the gauge dependence for the
obtained expression of nonlinear Landau damping rate was not considered.
In connection with above remarks it is evident that this problem
requires further investigation.

In this paper, we consider the above problem, using the approach
based on obtaining of generalized kinetic
equation for waves in quark-gluon plasma
developed by Kadomtsev, Silin, Tsytovich and others [11] in connection
with EMP as a basic method of
investigation of nonlinear processes in QGP.

We have shown that within the limits of nonlinear theory, developed for EMP
and {\it immediately} (i.e. without any crucial changes) applied to
investigation of nonlinear processes in QGP, nonlinear
Landau damping rate $\gamma^{l}({\bf k})$ defines not an inverse time of
damping at the expense of absorption (dissipation), but the inverse time of
spectral pumping of the energy of waves in the direction of a small wave numbers.
The inequality $\gamma^{l}(0)<0$ is a consequence of this fact, i.e.
${\bf k}=0$-mode, in contrast to [8], in this approximation increases.

The outline of the paper is as follows. In Sec.2,
we derive a system of self-consistent equations
in the covariant gauge
for regular and random parts of the distribution functions of  QGP
particles and gauge field is obtained. In Sec.3
the first order approximation of the color current is considered
and the correlation function of random oscillations is introduced. In
Secs.4 and 5 the second and the third orders approximation of the color
current are studied, and the terms leading in the coupling constant are
separated. In Sec.6 the generalized kinetic equation for waves in
QGP is derived. From there the kinetic equation describing
the nonlinear interaction of longitudinal oscillations is extracted in Sec.7.
In Sec.8 the physical mechanisms defining the nonlinear scattering
of waves by plasma particles are considered. In Sec.9 the estimate
of a value of the nonlinear Landau damping rate for longitudinal
eigenwaves is made. In Sec.10 the inaccuracies, which were made in [8]
in the process of computation of nonlinear Landau damping rate in the
temporal gauge, are
indicated, and its correct expression is written out. Comparison of this
expression with similar one, obtained in the covariant gauge is performed.
In Conclusion possible ways of further development of
the scrutinized theory are discussed.\\

{\bf 2. THE INITIAL EQUATIONS.
THE METHOD OF AVERAGING OVER
STATISTICAL ENSEMBLE}\\

We use metric $g^{\mu \nu} = diag(1,-1,-1,-1)$ and choose units such
that $c=k_{B}=1$. The gauge field potentials are $N_{c} \times N_{c}$-matrices
in a color space defined by $A_{\mu}=A_{\mu}^{a}t^{a}$ with $N_{c}^{2}-1$
hermitian generators of $SU(N_{c})$ group in the fundamental representation.
The field strength tensor $F_{\mu \nu}=F_{\mu \nu}^{a}t^{a}$ with
\begin{equation}
F_{\mu \nu}^{a} = \partial_\mu A_{\nu}^{a} - \partial_\nu A_{\mu}^{a}+
gf^{abc}A_{\mu}^{b}A_{\nu}^{c}
\label{eq:q}
\end{equation}
obeys the Yang-Mills (YM) equation in a covariant gauge
\begin{equation}
\partial_\mu F^{\mu \nu}(x) - ig[A_{\mu}(x),F^{\mu \nu}(x)] -
\xi^{-1} \partial^\nu \partial^\mu A_{\mu}(x) = -j^{\nu}(x),
\label{eq:w}
\end{equation}
where $ \xi$ is a gauge parameter. $j^{\nu}$ is the color current
\begin{equation}
j^{\nu} = gt^{a} \int d^{4}p \, p^{\nu}[
{\rm Sp} \,t^{a}(f_{q} - f_{\bar{q}})+{\rm Tr} \, (T^{a}f_{g})],
\label{eq:e}
\end{equation}
where $T^{a}$ are hermitian generators of $SU(N_{c})$ in
adjoint representation $((T^{a})^{bc}=-if^{abc}, {\rm Tr}(T^{a}T^{b})=N_{c}
\delta^{ab})$. We denote the trace over color indices in adjoint
representation as ${\rm Tr}$. Distribution functions of quarks $f_{q}$,
antiquarks $f_{\bar{q}}$, and gluons $f_{g}$ satisfy the semiclassical
kinetic equations (neglecting spin effects)
\begin{equation}
p^{\mu}{\cal D}_{\mu}f_{q,{\bar{q}}} \pm \frac{1}{2}gp^{\mu} \{ F_{\mu
\nu}, \frac{\partial f_{q,{\bar{q}}}}{\partial p_{\nu}} \} = 0,
\label{eq:r}
\end{equation}
\[
p^{\mu} \tilde{\cal D}_{\mu}f_{g}
+ \frac{1}{2}gp^{\mu} \{ {\cal F}_{\mu
\nu}, \frac{\partial f_{g}}{\partial p_{\nu}} \} = 0,
\]
where ${\cal D}_{\mu}$ and $\tilde{\cal D}_{\mu}$ are covariant
derivates which act as
\[
{\cal D}_{\mu} = \partial_{\mu} - ig[A_{\mu}(x), \cdot \, ],
\]
\[
 \tilde{\cal D}_{\mu} = \partial_{\mu} - ig[{\cal A}_{\mu}(x), \cdot \, ],
\]
[ , ] denotes commutator, $ \{ , \} $ denotes the anticommutator, and
${\cal A}_{\mu}$, ${\cal F}_{\mu \nu}$ are defined as ${\cal A}_{\mu}=
A_{\mu}^{a}T^{a}, {\cal F}_{\mu \nu} = F_{\mu \nu}^{a}T^{a}$. Upper sign
in the first equation (\ref{eq:r}) refers to quarks and lower one - to
antiquarks.

We begin with consideration of kinetic equations (\ref{eq:r}).
The distribution functions $f_{q,{\bar{q}}}$ and $f_{g}$
can be decomposed into two parts: regular and random (turbulent) ones
\begin{equation}
f_{s} = f_{s}^{R} + f_{s}^{T} \; , \; s=q, \bar{q}, g,
\label{eq:t}
\end{equation}
so that
\begin{equation}
\langle f_{s} \rangle = f_{s}^{R} \;, \; \langle f_{s}^{T} \rangle = 0,
\label{eq:y}
\end{equation}
where angular brackets $ \langle \cdot \rangle$ indicate a statistical
ensemble
of averaging. The initial values of parameters
which characterize the collective degree of a plasma freedom is
such statistical ensemble. For
almost linear collective motion to be considered below it may be
initial values of oscillation phases.

Further we set
\begin{equation}
A_{\mu} = A_{\mu}^{R} + A_{\mu}^{T} \;,\; \langle A_{\mu}^{T} \rangle = 0,
\label{eq:u}
\end{equation}
by definition.
For simplicity the regular part of the field $A_{\mu}^{R}$ will be
considered equal to zero.

Averaging the equation (\ref{eq:r}) over statistical ensemble, in view of
(\ref{eq:t})-(\ref{eq:u}), we obtain the equations for the regular part of the
distribution functions $f_{q, \bar{q}}^{R}$ and $f_{g}^{R}$
\[
p^{\mu} \partial_{\mu} f_{q, \bar{q}}^{R} = igp^{\mu}
\langle [A_{\mu}^{T},f_{q,{\bar{q}}}^{T}] \rangle \mp \frac{1}{2}gp^{\mu}
\langle \{ (F_{\mu \nu}^{T})_{L}, \frac{\partial f_{q, \bar{q}}^{T}}
{\partial p_{\nu}} \} \rangle \mp
\frac{1}{2}gp^{\mu}
\{ \langle (F_{\mu \nu}^{T})_{NL} \rangle ,
\frac{\partial f_{q, \bar{q}}^{R}}
{\partial p_{\nu}} \} \mp
\]
\begin{equation}
\mp \frac{1}{2}gp^{\mu}
\langle \{ (F_{\mu \nu}^{T})_{NL}, \frac{\partial f_{q, \bar{q}}^{T}}
{\partial p_{\nu}} \} \rangle ,
\label{eq:i}
\end{equation}
\[
p^{\mu} \partial_{\mu} f_{g}^{R} = igp^{\mu}
\langle [{\cal A}_{\mu}^{T},f_{g}^{T}] \rangle - \frac{1}{2}gp^{\mu}
\langle \{ ({\cal F}_{\mu \nu}^{T})_{L}, \frac{\partial f_{g}^{T}}
{\partial p_{\nu}} \} \rangle
- \frac{1}{2}gp^{\mu}
\{ \langle ({\cal F}_{\mu \nu}^{T})_{NL} \rangle, \frac{\partial f_{g}^{R}}
{\partial p_{\nu}} \} -
\]
\[
- \frac{1}{2}gp^{\mu}
\langle \{ ({\cal F}_{\mu \nu}^{T})_{NL}, \frac{\partial f_{g}^{T}}
{\partial p_{\nu}} \} \rangle .
\]
Here, indices $"L"$ and $"NL"$ denote the linear and nonlinear parts
with respect to field $A_{\mu}^{a}$ of the strength tensor (\ref{eq:q}).

Subtracting (\ref{eq:i}) from (\ref{eq:r}), we define the equations for
$f_{q, \bar{q}}^{T}$ and $f_{g}^{T}$
\[
p^{\mu} \partial_{\mu} f_{q, \bar{q}}^{T} = igp^{\mu}(
[A_{\mu}^{T},f_{q, \bar{q}}^{T}] -
\langle [A_{\mu}^{T},f_{q, \bar{q}}^{T}] \rangle)
\mp \frac{1}{2}gp^{\mu} \{ (F_{\mu \nu}^{T})_{L},
\frac{\partial f_{q, \bar{q}}^{R}}
{\partial p_{\nu}} \} \mp
\]
\[
\mp  \frac{1}{2}gp^{\mu}
( \{ (F_{\mu \nu}^{T})_{L}, \frac{\partial f_{q, \bar{q}}^{T}}
{\partial p_{\nu}} \} -
\langle \{ (F_{\mu \nu}^{T})_{L}, \frac{\partial f_{q, \bar{q}}^{T}}
{\partial p_{\nu}} \} \rangle) \mp
\frac{1}{2} g p^{\mu}
\{ (F_{\mu \nu}^{T})_{NL} -
\langle (F_{\mu \nu}^{T})_{NL} \rangle , \frac{\partial
f_{q, \bar{q}}^{R}}
{\partial p_{\nu}} \} \mp
\]
\begin{equation}
\mp \frac{1}{2}gp^{\mu}
( \{ (F_{\mu \nu}^{T})_{NL}, \frac{\partial
f_{q, \bar{q}}^{T}}
{\partial p_{\nu}} \} -
\langle \{ (F_{\mu \nu}^{T})_{NL}, \frac{\partial
f_{q, \bar{q}}^{T}}
{\partial p_{\nu}} \} \rangle ),
\label{eq:o}
\end{equation}
\[
p^{\mu} \partial_{\mu} f_{g}^{T} = igp^{\mu}(
[{\cal A}_{\mu}^{T},f_{g}^{T}] -
\langle [{\cal A}_{\mu}^{T},f_{g}^{T}] \rangle)
- \frac{1}{2}gp^{\mu} \{ ({\cal F}_{\mu \nu}^{T})_{L},
\frac{\partial f_{g}^{R}}
{\partial p_{\nu}} \} -
\]
\[
- \frac{1}{2} gp^{\mu}
( \{ ({\cal F}_{\mu \nu}^{T})_{L}, \frac{\partial f_{g}^{T}}
{ \partial p_{\nu}} \} -
\langle \{ ({\cal F}_{\mu \nu}^{T})_{L}, \frac{\partial f_{g}^{T}}
{\partial p_{\nu}} \} \rangle) -
\frac{1}{2} g p^{\mu}
\{ ({\cal F}_{\mu \nu}^{T})_{NL} -
\langle ({\cal F}_{\mu \nu}^{T})_{NL} \rangle , \frac{\partial
f_{g}^{R}}
{\partial p_{\nu}} \} -
\]
\[
- \frac{1}{2}gp^{\mu}
( \{ ({\cal F}_{\mu \nu}^{T})_{NL}, \frac{\partial
f_{g}^{T}}
{\partial p_{\nu}} \} -
\langle \{ ({\cal F}_{\mu \nu}^{T})_{NL}, \frac{\partial
f_{g}^{T}}
{\partial p_{\nu}} \} \rangle ).
\]
The system of equations (\ref{eq:i}) and (\ref{eq:o}) is suitable for
investigation of nonequilibrium processes in QGP such that the
excitation energy of waves is small quantity in relation to the total
energy of particles. In this case it can be used expansion in series
in power of oscillations amplitude of the random functions
$f_{s}^{T}$
\begin{equation}
f_{s}^{T}= \sum_{n=1}^{\infty} f_{s}^{T(n)} \;, \; s=q, \bar{q}, g,
\label{eq:p}
\end{equation}
where index $n$ depicts that $f_{s}^{T(n)}$ is proportional to
the $n$-th power of $A_{\mu}^{T}$. Substituting expansions (\ref{eq:p})
into (\ref{eq:o}), and collecting terms of the same order with respect to
$A_{\mu}^{T}$, we derive the system of equations
\begin{equation}
p^{\mu} \partial_{\mu} f_{q, \bar{q}}^{T(1)} = \mp \frac{1}{2}gp^{\mu}
\{ (F_{\mu \nu}^{T})_{L},
\frac{\partial f_{q, \bar{q}}^{R}}
{\partial p_{\nu}} \},
\label{eq:a}
\end{equation}
\begin{equation}
p^{\mu} \partial_{\mu} f_{q, \bar{q}}^{T(2)} = igp^{\mu}(
[A_{\mu}^{T},f_{q, \bar{q}}^{T(1)}] -
\langle [A_{\mu}^{T},f_{q, \bar{q}}^{T(1)}] \rangle) \mp
\label{eq:s}
\end{equation}
\[
\mp \frac{1}{2}gp^{\mu}
( \{ (F_{\mu \nu}^{T})_{L}, \frac{\partial f_{q, \bar{q}}^{T(1)}}
{\partial p_{\nu}} \} -
\langle \{ (F_{\mu \nu}^{T})_{L}, \frac{\partial f_{q, \bar{q}}^{T(1)}}
{\partial p_{\nu}} \} \rangle) \mp
\frac{1}{2} g p^{\mu}
\{ (F_{\mu \nu}^{T})_{NL} -
\langle (F_{\mu \nu}^{T})_{NL} \rangle , \frac{\partial
f_{q, \bar{q}}^{R}}
{\partial p_{\nu}} \} ,
\]
\begin{equation}
p^{\mu} \partial_{\mu} f_{q, \bar{q}}^{T(3)} = igp^{\mu}(
[A_{\mu}^{T},f_{q, \bar{q}}^{T(2)}] -
\langle [A_{\mu}^{T},f_{q, \bar{q}}^{T(2)}] \rangle) \mp
\frac{1}{2}gp^{\mu}
( \{ (F_{\mu \nu}^{T})_{L}, \frac{\partial f_{q, \bar{q}}^{T(2)}}
{\partial p_{\nu}} \} -
\label{eq:d}
\end{equation}
\[
- \langle \{ (F_{\mu \nu}^{T})_{L}, \frac{\partial f_{q, \bar{q}}^{T(2)}}
{\partial p_{\nu}} \} \rangle) \mp
\frac{1}{2}gp^{\mu}
( \{ (F_{\mu \nu}^{T})_{NL}, \frac{\partial
f_{q, \bar{q}}^{T(1)}}
{\partial p_{\nu}} \} -
\langle \{ (F_{\mu \nu}^{T})_{NL}, \frac{\partial
f_{q, \bar{q}}^{T(1)}}
{\partial p_{\nu}} \} \rangle ) \; etc..
\]
Similar equations are obtained for $f_{g}^{T(n)} \;, \; n=1,2,3, \ldots.$

The nonlinear color current is expressed as
\begin{equation}
j_{\mu}=j_{\mu}^{R} + j_{\mu}^{T} \;, \; \langle j_{\mu} \rangle =
j_{\mu}^{R} \;, \; j_{\mu}^{T}= \sum_{n=1}^{\infty} j_{\mu}^{T(n)},
\label{eq:f}
\end{equation}
where
\begin{equation}
j_{\mu}^{T(n)}= gt^{a} \int d^{4}p \, p_{\mu}[{\rm Sp} \, t^{a}(f_{q}^{T(n)}-
f_{\bar{q}}^{T(n)}) + {\rm Tr} \, (T^{a}f_{g}^{T(n)})].
\label{eq:g}
\end{equation}

Now we turn to the Yang-Mills equation (\ref{eq:w}), connecting the gauge
field with the color current. Averaging Eq. (\ref{eq:w}) and subtracting
the averaged equation from (\ref{eq:w}) in view of Eqs. (\ref{eq:j}) and
(\ref{eq:f}), we
find (for $A_{\mu}^{R}=0$)
\begin{equation}
\partial_{\mu}(F^{T \mu \nu})_{L} - \xi^{-1} \partial^{\nu}
\partial^{\mu}A_{\mu}^{T} + j^{T(1) \nu} = -j_{NL}^{T \nu} +
ig \partial_{\mu}([A^{T \mu},A^{T \nu}] - \langle [A^{T \mu},
A^{T \nu}] \rangle )+
\label{eq:h}
\end{equation}
\[
+ ig([A^{T}_{\mu},(F^{T \mu \nu})_{L}] - \langle [A_{\mu}^{T},
(F^{T \mu \nu})_{L}] \rangle ) +
g^{2}([A_{\mu}^{T},[A^{T \mu},A^{T \nu}]] -
\langle [A_{\mu}^{T},[A^{T \mu},A^{T \nu}]] \rangle ) .
\]
Here, in the left-hand side we collect all linear terms with
respect to $A_{\mu}^{T}$ and we denote: $j_{NL}^{T \nu} \equiv
j^{T(2) \nu} + j^{T(3) \nu} + \ldots.$ To account for
nonlinear interaction between waves
and particles in QGP (in first non-vanishing approximation
over the energy of waves), it is sufficiently
to restrict the consideration to the cubic terms with respect to
oscillation amplitude in expansion (\ref{eq:p}).

We introduce the following assumption. Eqs. (\ref{eq:i}) represent
the transport equations for averaged distribution functions.
The correlation functions in the right-hand side of these equations
have meaning of the collision terms of QGP particles with waves
and describe the influence of plasma waves to a background state.

We suppose that a characteristic time of nonlinear relaxation of
the oscillations is small quantity as compared with a time of relaxation
of the distribution particles $f_{s}^{R}$. Therefore
we neglect by variation of regular part of the distribution
functions with space and time, assuming that these functions are specified and
describe the global equilibrium in QGP
\begin{equation}
f_{q, \bar{q}}^{R} \equiv f_{q, \bar{q}}^{0} = 2
\frac{2N_{f} \theta(p_{0})}{(2 \pi)^{3}} \delta(p^{2})
\frac{1}{{\rm e}^{((pu) \mp \mu)/T} + 1}, \,
f_{g}^{R} \equiv f_{g}^{0} = 2
\frac{2 \theta(p_{0})}{(2 \pi)^{3}} \delta(p^{2})
\frac{1}{{\rm e}^{(pu)/T } - 1},
\label{eq:j}
\end{equation}
where $N_{f}$ - being the number of flavours for massless quarks,
$u_{\mu}$ is the four-velocity of the plasma at temperature $T$,
and $ \mu $ is the quark chemical potential.\\

{\bf 3. THE LINEAR APPROXIMATION.  THE CORRELATION FUNCTION
OF THE RANDOM OSCILLATIONS}\\

Let us turn our attention to obtaining of kinetic equation for
waves. The initial equation is Eq. (\ref{eq:h}). The left-hand side
of Eq. (\ref{eq:h}) contains a linear approximation of the color current,
explicit form of which is easily defined from Eq. (\ref{eq:a}).
We prefere to work in momentum space; the corresponding equations are
obtained by using
\[
A_{\mu}(x)= \int d^{4}k A_{\mu}(k) \, {\rm e}^{-i kx},
\]
and similar translations for $f^{T}_{q, \bar{q}}, \, f^{T}_{g}$. The result of
Fourier transformation for Eq. (\ref{eq:a}) is
\begin{equation}
f_{q, \bar{q}}^{T(1)}(k,p)= \mp g \frac{\chi^{\nu \lambda}(k,p)}
{pk + ip_{0} \epsilon} \frac{\partial f_{q, \bar{q}}^{0}}
{\partial p^{\lambda}} A_{\nu}(k) \, , \,
f_{g}^{T(1)}(k,p)= - g \frac{\chi^{\nu \lambda}(k,p)}
{pk + ip_{0} \epsilon} \frac{\partial f_{g}^{0}}
{\partial p^{\lambda}}{\cal A}_{\nu}(k),
\label{eq:k}
\end{equation}
\[
\epsilon \rightarrow +0.
\]
Here, $ \chi^{\nu \lambda}(k,p)=(pk)g^{\nu \lambda} - p^{\nu}k^{\lambda}$,
and the suffix $"T"$ for a gauge-field is omitted.
Substituting (\ref{eq:k}) into (\ref{eq:g}) (more precisely, in Fourier transformation
of (\ref{eq:g})) we define a well-known form [2, 12] of linear over field
approximation of the current
\begin{equation}
j^{T(1) \mu}(k)= \Pi^{\mu \nu}(k)A_{\nu}(k),
\label{eq:l}
\end{equation}
where
\[
\Pi^{\mu \nu}(k)=g^{2} \int \, d^{4}p \, \frac{p^{\mu}(p^{\nu}(k
\partial_{p})-(kp) \partial^{\nu}_{p}){\cal N}_{eq}}{pk + i p_{0}
\epsilon}
\]
is the high temperature polarization tensor, and
${\cal N}_{eq}= \frac{1}{2} (f_{q}^{0} + f_{\bar{q}}^{0}) + N_{c}
f_{g}^{0}$.

Further we rewrite Eq. (\ref{eq:h}) in the momentum space. Taking into account
(\ref{eq:l}), we obtain
\[
[k^{2}g^{\mu \nu} - (1+ \xi^{-1})k^{\mu}k^{\nu}- \Pi^{\mu \nu}(k)]
A^{b}_{\nu}(k)= j_{NL}^{Tb \mu}(k) +
\]
\begin{equation}
+ f^{bcd} \int S^{(I) \mu \nu \lambda}_{k,k_{1},k_{2}}
(A^{c}_{\nu}(k_{1})A_{\lambda}^{d}(k_{2})-
\langle A_{\nu}^{c}(k_{1})A_{\lambda}^{d}(k_{2}) \rangle)
\delta(k-k_{1}-k_{2}) dk_{1}dk_{2}+
\label{eq:z}
\end{equation}
\[
+ f^{bcf}f^{fde} \int \Sigma^{ \mu \nu \lambda \sigma}
_{k,k_{1},k_{2},k_{3}}
(A^{c}_{\nu}(k_{1})A_{\lambda}^{d}(k_{2})A_{\sigma}^{e}(k_{3})-
\langle A_{\nu}^{c}(k_{1})A_{\lambda}^{d}(k_{2})A_{\sigma}^{e}(k_{3})
\rangle)
\]
\[
\delta(k-k_{1}-k_{2}-k_{3}) \, dk_{1}dk_{2}dk_{3},
\]
where
\begin{equation}
S^{(I) \mu \nu \lambda}_{k,k_{1},k_{2}}=- ig(k^{\nu}
g^{\mu \lambda}+ k_{2}^{\nu}g^{\mu \lambda}- k_{2}^{\mu}g^{\nu \lambda}) \;,
\; \Sigma_{k,k_{1},k_{2},k_{3}}^{\mu \nu \lambda \sigma}=
g^{2}g^{\nu \lambda}g^{\mu \sigma}.
\label{eq:x}
\end{equation}

Let us multiply Eq. (\ref{eq:z}) by the complex conjugate
amplitude $A_{\mu}^{\ast a}(k^{\prime})$ and average it
\[
[k^{2}g^{\mu \nu} - (1+ \xi^{-1})k^{\mu}k^{\nu}- \Pi^{\mu \nu}(k)]
\langle A_{\mu}^{\ast a}(k^{\prime})A^{b}_{\nu}(k) \rangle
= \langle A_{\mu}^{\ast a}(k^{\prime})j_{NL}^{Tb \mu}(k) \rangle +
\]
\begin{equation}
+ f^{bcd} \int S^{(I) \mu \nu \lambda}_{k,k_{1},k_{2}}
\langle A_{\mu}^{\ast a}(k^{\prime})A_{\nu}^{c}(k_{1})
A_{\lambda}^{d}(k_{2}) \rangle
\delta(k-k_{1}-k_{2}) dk_{1}dk_{2}+
\label{eq:c}
\end{equation}
\[
+ f^{bcf}f^{fde} \int \Sigma^{\mu \nu \lambda \sigma}
_{k,k_{1},k_{2},k_{3}}
\langle A_{\mu}^{\ast a}(k^{\prime})A_{\nu}^{c}(k_{1})A_{\lambda}^{d}(k_{2})
A_{\sigma}^{e}(k_{3}) \rangle
\delta(k-k_{1}-k_{2}-k_{3}) dk_{1}dk_{2}dk_{3}.
\]

We introduce the correlation function of the random oscillations
\begin{equation}
I_{\mu \nu}^{ab}(k^{\prime},k)= \langle A_{\mu}^{\ast a}(k^{\prime})
A_{\nu}^{b}(k) \rangle.
\label{eq:v}
\end{equation}
In conditions of the stationary and homogeneous of QGP, i.e.
when the correlation function (\ref{eq:v}) in the coordinate representation
depends on the difference of coordinates and time $ \triangle x=x^{\prime}
- x$ only, we have
\begin{equation}
I_{\mu \nu}^{ab}(k^{\prime},k)=I_{\mu \nu}^{ab}(k^{\prime})
\delta (k^{\prime}-k).
\label{eq:b}
\end{equation}

By the effects of the nonlinear interaction of waves and particles,
the state of QGP becomes weakly inhomogeneous and weakly
nonstationary. The dependence of $I_{\mu \nu}^{ab}$ on $k^{\prime}-
k = \triangle k $ is no longer $ \delta $-shaped, i.e. $k^{\prime} \neq k$
and it "smeared" on $ \triangle k$, with $ \mid \triangle k / k \mid \ll 1$.

Let us introduce $I_{\mu \nu}^{ab}(k^{\prime},k)=I_{\mu \nu}^{ab}
(k, \triangle k)$, and insert the
correlation function
in the form
\[
I_{\mu \nu}^{ab}(k,x)= \int I_{\mu \nu}^{ab}(k, \triangle k)
{\rm e}^{- i \triangle kx} d \triangle k,
\]
slowly depending on $x$.
In Eq. (\ref{eq:c}) we make the change $k \rightleftharpoons
k^{\prime} \;, \; a \rightleftharpoons b $, complex conjugate
and subtract obtained equation from Eq. (\ref{eq:c}), beforehand
expanding of the polarization tenzor into Hermitian and anti-Hermitian
parts
\[
\Pi^{\nu \sigma}(k)= \Pi^{H \nu \sigma}(k) + \Pi^{A \nu \sigma}(k) \;,
\; \Pi^{H \nu \sigma}(k)= \Pi^{\ast H \sigma \nu}(k) \;, \;
\Pi^{A \nu \sigma}(k)= - \Pi^{\ast A \sigma \nu}(k).
\]

By using the definition (\ref{eq:v}), we obtain the result
\[
[(k^{2}g^{\mu \nu}-(1+ \xi^{-1})k^{\mu}k^{\nu}) -
(k^{\prime 2}g^{\mu \nu}-(1+ \xi^{-1})k^{\prime \mu}k^{\prime \nu})-
( \Pi^{H \mu \nu}(k)- \Pi^{H \mu \nu}(k^{\prime}))] I_{\mu \nu}^{ab}
(k^{\prime},k)-
\]
\[
-[ \Pi^{A \mu \nu}(k)+ \Pi^{A \mu \nu}(k^{\prime})] I_{\mu \nu}^{ab}
(k^{\prime},k)=
\langle A_{\mu}^{\ast a}(k^{\prime})j_{NL}^{Tb \mu}(k) \rangle -
\langle A_{\mu}^{b}(k)j_{NL}^{\ast Ta \mu}(k^{\prime}) \rangle +
\]
\[
+ f^{bcd} \int S_{k,k_{1},k_{2}}^{(I) \mu \nu \lambda} \langle
A_{\mu}^{\ast a}(k^{\prime})A_{\nu}^{c}(k_{1})A_{\lambda}^{d}
(k_{2}) \rangle \delta (k-k_{1}-k_{2}) dk_{1} dk_{2}-
\]
\begin{equation}
- f^{acd} \int S_{k^{\prime},k_{1},k_{2}}^{\ast (I) \mu \nu \lambda}
\langle A_{\mu}^{b}(k)A_{\nu}^{\ast c}(k_{1})A_{\lambda}^{\ast d}
(k_{2}) \rangle \delta (k^{\prime} - k_{1} - k_{2}) dk_{1} dk_{2}+
\label{eq:n}
\end{equation}
\[
+f^{bcf}f^{fde} \int \Sigma_{k,k_{1},k_{2},k_{3}}^{\mu \nu \lambda
\sigma} \langle A_{\mu}^{\ast a}(k^{\prime})A_{\nu}^{c}(k_{1})
A_{\lambda}^{d}(k_{2})A_{\sigma}^{e}(k_{3}) \rangle
\delta (k-k_{1}-k_{2}-k_{3}) dk_{1}dk_{2}dk_{3}-
\]
\[
-f^{acf}f^{fde} \int \Sigma_{k^{\prime},k_{1},k_{2},k_{3}}
^{\ast \mu \nu \lambda
\sigma} \langle A_{\mu}^{b}(k)A_{\nu}^{\ast c}(k_{1})
A_{\lambda}^{\ast d}(k_{2})A_{\sigma}^{\ast e}(k_{3}) \rangle
\delta (k^{\prime}-k_{1}-k_{2}-k_{3}) dk_{1}dk_{2}dk_{3}.
\]

We assume that anti-Hermitian part of $ \Pi^{A}$ is small in comparison
with $ \Pi^{H}$ and it is a value of the same smallness order, as the
nonlinear terms in the right-hand side of Eq. (\ref{eq:n}). Therefore it can be suggested
that $ \Pi^{A \nu \sigma}(k) \simeq \Pi^{A \nu \sigma}(k^{\prime})$,
and the term with $ \Pi^{A}$ can be rearranged to the right-hand
side of Eq. (\ref{eq:n}).The remaining terms in the left-hand side of Eq.
(\ref{eq:n})
we expanded in a series in powers of $ \triangle k$ to first smallness order.
Multiplying obtained equation by ${\rm e}^{- i \triangle kx}$
and integrating over $ \triangle k$ with regard to
\[
\int \triangle k_{\lambda} \, I_{\mu \nu}^{ab}(k, \triangle k)
\, {\rm e}^{- i \triangle kx} d \triangle k=
i \frac{\partial I_{\mu \nu}^{ab}(k,x)}{\partial x^{\lambda}},
\]
we obtain finally
\[
\frac{\partial}{\partial k_{\lambda}}[k^{2}g^{\mu \nu}
-(1+ \xi^{-1})k^{\mu}k^{\nu} -
\Pi^{H \mu \nu}(k)] \frac{\partial I_{\mu \nu}^{ab}}{\partial x^{\lambda}}
=2 i \Pi^{A \mu \nu}I_{\mu \nu}^{ab}-
\]
\[
- i \int dk^{\prime} \{
\langle A_{\mu}^{\ast a}(k^{\prime})j_{NL}^{Tb \mu}(k) \rangle -
\langle A_{\mu}^{b}(k)j_{NL}^{\ast Ta \mu}(k^{\prime}) \rangle \} -
\]
\[
- i \{ f^{bcd} \int \, dk^{\prime}dk_{1}dk_{2} \,
S_{k,k_{1},k_{2}}^{(I) \mu \nu \lambda} \langle
A_{\mu}^{\ast a}(k^{\prime})A_{\nu}^{c}(k_{1})A_{\lambda}^{d}
(k_{2}) \rangle \delta (k-k_{1}-k_{2}) -
\]
\begin{equation}
- f^{acd} \int \, dk^{\prime}dk_{1}dk_{2} \,
S_{k^{\prime},k_{1},k_{2}}^{\ast (I) \mu \nu \lambda}
\langle
A_{\mu}^{b}(k)A_{\nu}^{\ast c}(k_{1})A_{\lambda}^{\ast d}
(k_{2}) \rangle \delta (k^{\prime}-k_{1}-k_{2}) \} -
\label{eq:m}
\end{equation}
\[
- i \{ f^{bcf}f^{fde} \int \Sigma_{k,k_{1},k_{2},k_{3}}^{\mu \nu \lambda
\sigma} \langle A_{\mu}^{\ast a}(k^{\prime})A_{\nu}^{c}(k_{1})
A_{\lambda}^{d}(k_{2})A_{\sigma}^{e}(k_{3}) \rangle
\delta (k-k_{1}-k_{2}-k_{3}) dk_{1}dk_{2}dk_{3}-
\]
\[
-f^{acf}f^{fde} \int \Sigma_{k^{\prime},k_{1},k_{2},k_{3}}
^{\ast \mu \nu \lambda
\sigma} \langle A_{\mu}^{b}(k)A_{\nu}^{\ast c}(k_{1})
A_{\lambda}^{\ast d}(k_{2})A_{\sigma}^{\ast e}(k_{3}) \rangle
\delta (k^{\prime}-k_{1}-k_{2}-k_{3}) dk_{1}dk_{2}dk_{3} \},
\]
where $j_{NL}^{Ta \mu}(k)=j^{T(2)a \mu}(k) +
j^{T(3)a \mu}(k)$.

We made several remarks relative to obtained Eq. (\ref{eq:m}). The term
with $\Pi^{A}$ introducing in the right-hand side of Eq. (\ref{eq:m}) corresponds
to linear Landau damping. However, as was shown by Heinz and
Siemens [9], linear Landau damping for waves in QGP is absent
and hence this term vanishes.

As it will be shown below, the terms with $\Sigma$ make no
contribute to processes of the nonlinear interaction of waves with particles
of QGP, and hereafter they will be dropped.\\

{\bf 4. THE SECOND APPROXIMATION OF THE COLOR CURRENT}\\

Now we conserned with computation of the nonlinear corrections
to the current in the right-hand side of basic Eq. (\ref{eq:m}).
In this section we consider the second order of $j^{T(2) \mu}(k)$.

At first we define $f_{q, \bar{q}}^{T(2)}$. We carry out the Fourier
transformation of Eq. (\ref{eq:s})
\[
-i(pk) f_{q, \bar{q}}^{T(2)}(k,p)= gp^{\mu} \int \, \Big{\{} i(
[A_{\mu}(k_{1}),f_{q, \bar{q}}^{T(1)}(k_{2},p)]-
\langle [A_{\mu}(k_{1}),f_{q, \bar{q}}^{T(1)}(k_{2},p)] \rangle) \mp
\]
\[
\mp \frac{1}{2}( \{ (F_{\mu \nu})_{L}(k_{1}), \frac{\partial
f_{q, \bar{q}}^{T(1)}(k_{2},p)}{\partial p_{\nu}} \}-
\langle \{ (F_{\mu \nu})_{L}(k_{1}), \frac{\partial
f_{q, \bar{q}}^{T(1)}(k_{2},p)}{\partial p_{\nu}} \} \rangle ) \pm
\]
\[
\pm ig([A_{\mu}(k_{1}),A_{\nu}(k_{2})] -
\langle [A_{\mu}(k_{1}),A_{\nu}(k_{2})] \rangle)
\frac{\partial f_{q, \bar{q}}^{0}}{\partial p_{\nu}} \Big{\}} \delta (k - k_{1} -
k_{2}) dk_{1}dk_{2}.
\]

Substituting the obtained $f_{q, \bar{q}}^{T(1)}$ from (\ref{eq:k}) into the
last equation and collecting similar terms, we obtain
\begin{equation}
f_{q, \bar{q}}^{T(2)}= \mp g^{2} \frac{[t^{b},t^{c}] p^{\nu}p^{\lambda}}
{pk + ip_{0} \epsilon} \int \, \frac{(k_{2} \partial_{p} f_{q, \bar{q}}
^{0})}{pk_{2} + ip_{0} \epsilon}
(A_{\nu}^{b}(k_{1})A_{\lambda}^{c}(k_{2})
- \langle A_{\nu}^{b}(k_{1})A_{\lambda}^{c}(k_{2}) \rangle)
\delta (k - k_{1} - k_{2}) dk_{1}dk_{2} +
\label{eq:qq}
\end{equation}
\[
+ \frac{g^{2}}{2} \frac{ \{ t^{b},t^{c} \} }
{pk + ip_{0} \epsilon} \int \, \chi^{\nu \lambda}(k_{1},p)
\frac{\partial}{\partial p^{\lambda}} \left(
\frac{\chi ^{\sigma  \rho}(k_{2},p)}{pk_{2} + ip_{0} \epsilon} \,
\frac{\partial f_{q, \bar{q}}^{0}}{\partial p^{\rho}} \right)
(A_{\nu}^{b}(k_{1})A_{\sigma}^{c}(k_{2})
- \langle A_{\nu}^{b}(k_{1})A_{\sigma}^{c}(k_{2}) \rangle )
\]
\[
\delta (k - k_{1} - k_{2}) \, dk_{1}dk_{2}.
\]
The expression for $f_{g}^{T(2)}(k,p)$ is obtained from (\ref{eq:qq})
by choosing upper sign and replacements $f_{q}^{0} \rightarrow f_{g}^{0},
t^{a} \rightarrow T^{a}$. Substituting obtained expressions $f^{T(2)}_{s},
s=q, \bar{q},g$ into (\ref{eq:g}) (for $n=2$), we find required current correction
\begin{equation}
j^{T(2)a \mu}(k)= -ig^{3}f^{abc} \int \, d^{4}p \,
\frac{p^{\mu}p^{\nu}p^{\lambda}}
{pk + ip_{0} \epsilon} \, \frac{(k_{2} \partial_{p} {\cal N}_{eq}
)}{pk_{2} + ip_{0} \epsilon} \,
(A_{\nu}^{b}(k_{1})A_{\lambda}^{c}(k_{2})
- \langle A_{\nu}^{b}(k_{1})A_{\lambda}^{c}(k_{2}) \rangle)
\label{eq:ww}
\end{equation}
\[
\delta (k - k_{1} - k_{2}) \, dk_{1}dk_{2} +
\]
\[
+ \frac{g^{3}}{4}d^{abc} \int \, d^{4}p \, \frac{p^{\mu} \chi^{\nu \lambda}
(k_{1},p)}{pk + ip_{0} \epsilon} \frac{\partial}{\partial
p^{\lambda}} \left(
\frac{\chi ^{\sigma  \rho}(k_{2},p)}{pk_{2} + ip_{0} \epsilon} \,
\frac{\partial (f_{q}^{0}-f_{\bar{q}}^{0})}{\partial p^{\rho}} \right)
(A_{\nu}^{b}(k_{1})A_{\sigma}^{c}(k_{2})-
\langle A_{\nu}^{b}(k_{1})A_{\sigma}^{c}(k_{2}) \rangle)
\]
\[
\delta (k - k_{1} - k_{2}) \, dk_{1}dk_{2}.
\]
The contribution of gluons to the expression with symmetric structure
constant $d^{abc}$ here drops out. This is connected with
the fact that in calculation of trace of anti-commutators we have:
${\rm Sp} \, t^{a}
\{ t^{b},t^{c} \}= \frac{1}{2}d^{abc}$ -for quarks and antiquarks, and
${\rm Tr} \, T^{a} \{ T^{b},T^{c} \} =0$ - for gluons. The symmetry of
contributions can be restored if we note that besides usual gluon
current $j_{g}^{\mu}(x)= gt^{a} \int \, d^{4}p \, p^{\mu} {\rm Tr}
(T^{a}f_{g}(x,p))$, the kinetic equation for gluons admits
a covariant conserving quantity
\begin{equation}
\lambda gt^{a} \int \, d^{4}p \, {\rm Tr}({\cal P}^{a}f_{g}(x,p)),
\label{eq:ee}
\end{equation}
where $({\cal P}^{a})^{bc}=d^{abc}$ and $ \lambda$ is a certain arbitrary
constant. The covariant continuity of (\ref{eq:ee}) is evident from the identity:
$[{\cal P}^{a},T^{b}]=if^{abc} {\cal P}^{c}$. On addition of (\ref{eq:ee}) to
(\ref{eq:e}) we have contributions to the nonlinear current corrections only.
Adding (\ref{eq:ee}) to the second current iteration (\ref{eq:g}) and taking into
account the equality
\[
{\rm Sp} \, {\cal P}^{a} \{ T^{b},T^{c} \}= N_{c}d^{abc},
\]
instead of (\ref{eq:ww}) we derive more general expression for $j^{T(2)}$
\begin{equation}
j^{T(2)a \mu}(k)= \int \, S^{abc \mu \nu \lambda}_{k,k_{1},k_{2}}
(A^{b}_{\nu}(k_{1})A_{\lambda}^{c}(k_{2})-
\langle A^{b}_{\nu}(k_{1})A_{\lambda}^{c}(k_{2}) \rangle)
\delta (k - k_{1} - k_{2}) dk_{1}dk_{2},
\label{eq:rr}
\end{equation}
where $\, S_{k,k_{1},k_{2}}^{abc \mu \nu \lambda}=
f^{abc}S_{k,k_{1},k_{2}}^{(II) \mu \nu \lambda}+
d^{abc}S_{k,k_{1},k_{2}}^{(III) \mu \nu \lambda}$,
\begin{equation}
S_{k,k_{1},k_{2}}^{(II) \mu \nu \lambda}=
-ig^{3} \int \, d^{4}p \, \frac{p^{\mu}p^{\nu}p^{\lambda}}
{pk + ip_{0} \epsilon} \, \frac {(k_{2} \partial_{p} {\cal N}_{eq})}
{pk_{2} + ip_{0} \epsilon},
\label{eq:tt}
\end{equation}
\begin{equation}
S_{k,k_{1},k_{2}}^{(III) \mu \nu \lambda}=
\frac{g^{3}}{2} \int \, d^{4}p \, \frac{p^{\mu} \chi^{\nu \sigma}(k_{1},p)}
{pk + ip_{0} \epsilon} \, \frac{\partial}{\partial p^{\sigma}}
\left( \frac {\chi^{\lambda \rho}(k_{2},p)}
{pk_{2} + ip_{0} \epsilon} \, \frac{{\partial \cal M}_{eq}}
{\partial p^{\rho}} \right),
\label{eq:yy}
\end{equation}
\[
{\cal M}_{eq}= \frac{1}{2}(f_{q}^{0} - f_{\bar{q}}^{0}) +
\lambda N_{c}f_{g}^{0}.
\]

The tensor structure of $S^{(III) \mu \nu \lambda}_{k,k_{1},k_{2}}$
exactly coincides with appropriate expression obtained in
calculation of $j^{T(2) \mu}$ in electromagnetic plasmas [11],
and hence part of current with $d^{abc}$ has
a meaning of Abelian part of the color current $j^{T(2)a \mu}$.
The term with $S_{k,k_{1},k_{2}}^{(II) \mu \nu \lambda}$
is purely non-Abelian, i.e. it has no Abelian counterpart.

Let us estimate orders of $S^{(II)}$ and $S^{(III)}$. Following
usual terminology [1], we call an energy or a momentum "soft"
when it is of order $gT$, and "hard" when it is of order $T$. We
will be considered, as in Ref. [5], that collective excitations
carrying soft momenta, i.e. $k \sim gT$, and plasma particles
have the typical hard energies: $p \sim T$. On this basis, we have the
following estimate for $S^{(II)}$
\begin{equation}
S_{k,k_{1},k_{2}}^{(II) \mu \nu \lambda} \sim g^{2}T.
\label{eq:ii}
\end{equation}
Here, we considered that by virtue of the definitions (\ref{eq:j}):
${\cal N}_{eq} \sim 1/T^{2}$.

In expression (\ref{eq:yy}) the integral of energy with gluon distribution
function is infrared divergent. In a similar manner [8], we regulate it
by introducing an electric mass cut-off of order $gT$, and only take
the leading term in $g$. Then it can be found that in (\ref{eq:yy}), the
part related to the gluon distribution function is of order $g^{2}T
(g \ln g)$ and the other part related to the quark and antiquark
distribution functions is of order $g^{3}T$, i.e.
\[
S_{k,k_{1},k_{2}}^{(III) \mu \nu \lambda} \sim g^{2}T(g \ln g)+
g^{3}T.
\]

Hence $S^{(II)}$, which is purely non-Abelian, is of lower order in
the coupling constant than $S^{(III)}$, which has an Abelian
counterpart. This fact was first seen in Ref. [8].\\

{\bf 5. THE THIRD APPROXIMATION OF THE COLOR CURRENT}\\

Now we calculate the nonlinear correction of $j^{T(3) \mu}$. We carry out the
Fourier transformation of Eq. (\ref{eq:d}) and taking into account (\ref{eq:k}),
(\ref{eq:qq}), after cumbersome transformations, we define
\begin{center}
\[
f_{q, \bar{q}}^{T(3)}(k,p)= \int \, \bigg[
\mp g^{3} \frac{f^{bcf} f^{cde} t^{f}}{pk + ip_{0} \epsilon}
\, \frac{p^{\nu} p^{\lambda} p^{\sigma}}{p(k_1 + k_2) + ip_{0} \epsilon}
\, \frac{(k_{2} \partial_{p} f_{q, \bar{q}}^{0})}{pk_{2} + ip_{0} \epsilon}-
\]
\end{center}
\[
- \frac{ig^3}{2} \frac{[t^{b}, \{ t^{d},t^{e} \} ]}{pk + ip_{0} \epsilon}
\, \frac{p^{\nu} \chi^{\lambda \alpha}(k_{1},p)}{p(k_{1} + k_{2})
+ ip_{0} \epsilon}
\, \frac{\partial}{\partial p^{\alpha}} \Big(
\frac{\chi^{\sigma \rho}(k_2,p)}{pk_2 + ip_{0} \epsilon}
\frac{\partial f_{q, \bar{q}}^0}
{\partial p^{\rho}} \Big) +
\]
\begin{equation}
+ \frac{ig^3}{2} \frac{f^{cde} \{ t^{b},t^{c} \}}{pk + ip_{0} \epsilon}
\chi^{\nu \tau}(k_{3},p) \frac{\partial}{\partial p^{\tau}} \Big(
\frac{p^{\lambda}p^{\sigma}}{p(k_{1} +k_{2}) + ip_{0} \epsilon}
\, \frac{(k_{2} \partial_{p}f_{q, \bar{q}}^{0})}{pk_{2} + ip_{0} \epsilon}
\Big) \mp
\label{eq:oo}
\end{equation}
\[
\mp \frac{g^{3}}{4} \frac{\{ t^{b}, \{ t^{d},t^{e} \} \}}
{pk + ip_{0} \epsilon} \chi^{\nu \tau}(k_{3},p)
\frac{\partial}{\partial p^{\tau}} \Big(
\frac{\chi^{\lambda \alpha}(k_{1},p)}{p(k_{1} + k_{2}) + ip_{0} \epsilon}
\, \frac{\partial}{\partial p^{\alpha}} \Big(
\frac{\chi^{\sigma \rho}(k_{2},p)}{pk_{2} + ip_{0} \epsilon}
\frac{\partial f_{q, \bar{q}}^{0}}{\partial p^{\rho}} \Big) \Big) \bigg]
\]
\[
(A_{\nu}^{b}(k_{3})A_{\lambda}^{d}(k_{1})A_{\sigma}^{e}(k_{2})-
A_{\nu}^{b}(k_{3}) \langle A_{\lambda}^{d}(k_{1})A_{\sigma}^{e}(k_{2}) \rangle -
\langle A_{\nu}^{b}(k_{3})A_{\lambda}^{d}(k_{1})A_{\sigma}^{e}(k_{2}) \rangle)
\]
\[
\delta (k - k_{1} - k_{2} - k_{3}) \, dk_{1}dk_{2}dk_{3} +
\]
\[
+ \frac{ig^{3}}{2} \int \,
\frac{f^{cde} \{ t^{b},t^{c} \}}{pk + ip_{0} \epsilon}
p^{\lambda}
\frac{\partial}{\partial p^{\sigma}} \Big(
\frac{\chi^{\nu \rho}(k_{3},p)}{pk_{3} + ip_{0} \epsilon}
\frac{\partial f_{q, \bar{q}}^{0}}{\partial p^{\rho}} \Big)
(A_{\nu}^{b}(k_{3})A_{\lambda}^{d}(k_{1})A_{\sigma}^{e}(k_{2})-
\langle A_{\nu}^{b}(k_{3})A_{\lambda}^{d}(k_{1})A_{\sigma}^{e}(k_{2}) \rangle )
\]
\[
\delta (k - k_1 - k_2 - k_3) \, dk_{1} dk_{2} dk_{3}.
\]
The gluon part of $f_{g}^{T(3)}(k,p)$ is obtained from (\ref{eq:oo})
by replacements $f_{q}^{0} \rightarrow f_{g}^{0}$
and $t^{a} \rightarrow T^{a}$. Substituting obtained expressions
for $f_{q, \bar{q}}^{T(3)}$ and $f_{g}^{T(3)}$ into (\ref{eq:g}) (with regard
to additional current (\ref{eq:ee})) and taking into account the equalities
\begin{equation}
{\rm Tr} \, ( \{ T^{a},T^{b} \} \{ T^{d},T^{e} \} )=
N_{c}d^{abc}d^{cde} + 4 \delta^{ab} \delta^{ed} + 2 \delta^{ad}
\delta^{eb} + 2 \delta^{ae} \delta^{bd},
\label{eq:pp}
\end{equation}
\[
{\rm Tr} \, ( \{ {\cal P}^{a},T^{b} \} \{ T^{d},T^{e} \} )=0,
\]
we find the required form of $j^{T(3)a \mu}$
\[
j^{T(3)a \mu}(k)= \int \, \Sigma_{k,k_{1},k_{2},k_{3}}^{abde \mu
\nu \lambda \sigma}(
A_{\nu}^{b}(k_{3})A_{\lambda}^{d}(k_{1})A_{\sigma}^{e}(k_{2})-
A_{\nu}^{b}(k_{3}) \langle A_{\lambda}^{d}(k_{1})
A_{\sigma}^{e}(k_{2}) \rangle -
\]
\begin{equation}
- \langle A_{\nu}^{b}(k_{3})A_{\lambda}^{d}(k_{1})A_{\sigma}^{e}(k_{2}) \rangle )
\delta (k - k_{1} - k_{2} - k_{3}) \, dk_{1}dk_{2}dk_{3} +
\label{eq:aa}
\end{equation}
\[
+ d^{abc}f^{cde} \int \, R_{k,k_{1},k_{2},k_{3}}^{ \mu
\nu \lambda \sigma}(
A_{\nu}^{b}(k_{3})A_{\lambda}^{d}(k_{1})A_{\sigma}^{e}(k_{2})-
\langle A_{\nu}^{b}(k_{3})A_{\lambda}^{d}(k_{1})A_{\sigma}^{e}(k_{2}) \rangle )
\]
\[
\delta (k - k_{1} - k_{2} - k_{3}) \, dk_{1}dk_{2}dk_{3}.
\]
Here,
\[
\Sigma_{k,k_{1},k_{2},k_{3}}^{abde \mu \nu \lambda \sigma}=
f^{abc}f^{cde} \Sigma_{k,k_{1},k_{2},k_{3}}^{(I) \mu \nu \lambda
\sigma}+
f^{abc}d^{cde} \Sigma_{k,k_{1},k_{2},k_{3}}^{(II) \mu \nu \lambda
\sigma}+
\delta^{ab} \delta^{de} \Sigma_{k,k_{1},k_{2},k_{3}}^{(III) \mu \nu \lambda
\sigma}+
d^{abc}f^{cde} \Sigma_{k,k_{1},k_{2},k_{3}}^{(IV) \mu \nu \lambda
\sigma} +
\]
\[
+ d^{abc}d^{cde} \Sigma_{k,k_{1},k_{2},k_{3}}^{(V) \mu \nu \lambda
\sigma}+
( \delta^{ab} \delta^{de} + \delta^{ad} \delta^{be} + \delta^{ae} \delta^{db})
\Sigma_{k,k_{1},k_{2},k_{3}}^{(VI) \mu \nu \lambda \sigma},
\]
\begin{equation}
\Sigma_{k,k_{1},k_{2},k_{3}}^{(I) \mu \nu \lambda \sigma} =
-g^{4} \int \, d^{4}p \,
\frac{p^{\mu}p^{\nu}p^{\lambda}p^{\sigma}}{pk + ip_{0} \epsilon}
\, \frac{1}{p(k_{1} +k_{2}) + ip_{0} \epsilon}
\, \frac{(k_{2} \partial_p {\cal N}_{eq})}{pk_{2} + ip_{0} \epsilon},
\label{eq:ss}
\end{equation}
\begin{equation}
\Sigma_{k,k_{1},k_{2},k_{3}}^{(III) \mu \nu \lambda \sigma}=
- \frac{g^{4}}{2N_{c}} \int \, d^{4}p \frac{p^{\mu} \chi^{\nu \tau}(k_{3},p)}
{pk + ip_{0} \epsilon}
\frac{\partial}{\partial p^{\tau}} \Big(
\frac{\chi^{\lambda \alpha}(k_{1},p)}{p(k_{1} + k_{2}) + ip_{0} \epsilon}
\, \frac{\partial}{\partial p^{\alpha}} \Big(
\frac{\chi^{\sigma \rho}(k_{2},p)}{pk_{2} + ip_{0} \epsilon}
\frac{\partial {\cal N}_{eq}}{\partial p^{\rho}} \Big) \Big),
\label{eq:dd}
\end{equation}
\[
\Sigma_{k,k_{1},k_{2},k_{3}}^{(V) \mu \nu \lambda \sigma}=
\frac{N_{c}}{2} \Sigma_{k,k_{1},k_{2},k_{3}}^{(III) \mu \nu \lambda \sigma}.
\]
The expression for $ \Sigma^{(VI)}$ is obtained from (\ref{eq:dd}) by exception
of quark and antiquark contributions. The availability of the term with
$ \Sigma^{(VI)}$ is reflection of more complicated color structure
of the gluon kinetic equation in comparison with quark and antiquark equations.

The terms with $ \Sigma^{(II)}, \Sigma^{(IV)}$ and $R$ are defined as
the interference of Abelian and non-Abelian contributions. In the case
of isotropic, homogeneous and colorless plasma, the correlation
function (\ref{eq:v}) is proportional to unit matrix in color space, i.e.
plasma oscillations are degenerate in color association. This leads
to the fact that the coefficients standing before these interference terms,
supporting the color indicies, are vanish in kinetic equation for waves and
therefore their explicit form is not given here. With the availability,
e.g., of external color field only, when degeneration is removed
(the correlation function (\ref{eq:vv}) becomes nontrivial matrix in color
space), the interference of Abelian and non-Abelian contributions will
be presented.

At the end of this section we estimate the order of $ \Sigma^{(I)}$ and
$ \Sigma^{(III)}$. It follows from the expression (\ref{eq:ss}) that
\begin{equation}
\Sigma_{k,k_{1},k_{2},k_{3}}^{(I) \mu \nu \lambda \sigma}
\sim g^{2}.
\label{eq:ff}
\end{equation}
Cutting off, as in the previous section, integration limit for the gluon
distribution function, we find
\[
\Sigma^{(III)} \sim \Sigma^{(V)} \sim g^{3} + g^{4} \;,
\; \Sigma^{(VI)} \sim g^{3}.
\]
By this means, purely non-Abelian contribution of $ \Sigma^{(I)}$ is
of lower order in the coupling constant than Abelian - $ \Sigma^{(III)},
\Sigma^{(V)}$ and $ \Sigma^{(VI)}$.\\

{\bf 6. THE GENERALIZED KINETIC EQUATION FOR WAVES}\\

Let us consider the initial equation for waves (\ref{eq:m}). Substituting
obtained nonlinear corrections of the induced current
by field (\ref{eq:rr}) and (\ref{eq:aa}) into this equation, and considering the terms of leading order in $g$ only,
we obtain
\[
\frac{\partial}{\partial k_{\lambda}}[k^{2}g^{\mu \nu} - (1 + \xi^{-1})
k^{\mu}k^{\nu} - \Pi^{H \mu \nu}(k)] \frac{\partial I_{\mu \nu}^{ab}}
{\partial x^{\lambda}}=
\]
\[
= -i \int \, dk^{\prime} \{
f^{bcd}S_{k,k_{1},k_{2}}^{\mu \nu \lambda} \langle
A_{\mu}^{\ast a}(k^{\prime})A_{\nu}^{c}(k_{1})A_{\lambda}^{d}(k_{2})
\rangle dk_{1} dk_{2} \delta(k - k_{1} - k_{2})-
\]
\[
-f^{acd}S_{k^{\prime},k_{1},k_{2}}^{\ast \mu \nu \lambda} \langle
A_{\mu}^{b}(k)A_{\nu}^{\ast c}(k_{1})A_{\lambda}^{\ast d}(k_{2})
\rangle dk_{1} dk_{2} \delta(k^{\prime} - k_{1} - k_{2})+
\]
\[
+ f^{bcf}f^{fde} \Sigma_{k,k_{1},k_{2},k_{3}}^{(I) \mu \nu \lambda \sigma}
( \langle A_{\mu}^{\ast a}(k^{\prime})A_{\nu}^{c}(k_{3})A_{\lambda}^{d}(k_{1})
A_{\sigma}^{e}(k_{2}) \rangle -
\langle A_{\mu}^{\ast a}(k^{\prime})A_{\nu}^{c}(k_{3}) \rangle
\langle A_{\lambda}^{d}(k_{1})A_{\sigma}^{e}(k_{2}) \rangle )
\]
\begin{equation}
dk_{1} dk_{2} dk_{3} \delta (k - k_{1} - k_{2} - k_{3})-
\label{eq:gg}
\end{equation}
\[
- f^{acf}f^{fde} \Sigma_{k^{\prime},k_{1},k_{2},k_{3}}^{\ast
(I) \mu \nu \lambda \sigma}
( \langle A_{\mu}^{b}(k)A_{\nu}^{\ast c}(k_{3})A_{\lambda}^{\ast d}(k_{1})
A_{\sigma}^{\ast e}(k_{2}) \rangle -
\langle A_{\mu}^{b}(k^{\prime})A_{\nu}^{\ast c}(k_{3}) \rangle
\langle A_{\lambda}^{\ast d}(k_{1})A_{\sigma}^{\ast e}(k_{2}) \rangle )
\]
\[
dk_{1} dk_{2} dk_{3} \delta (k^{\prime} - k_{1} - k_{2} - k_{3}) \}.
\]

Here,
$S_{k,k_{1},k_{2}}^{\mu \nu \lambda} \equiv S_{k,k_{1},k_{2}}^{(I)
\mu \nu \lambda} + S_{k,k_{1},k_{2}}^{(II) \mu \nu \lambda}.$

By virtue of weak nonlinearity, oscillations phases of a field weakly correlate
among themselves. Therefore mean value of four random quantities can
be approximately divided into product of mean values of two
fields. For mean value of three fields this decomposition vanishes,
and it should be considered a weak correlation of fields.
For this purpose we use the nonlinear equation of a field (\ref{eq:z}),
taking into account in the right-hand side of (\ref{eq:z}) the terms of the second
order in  $A$ only
\[
[k^{2}g^{\mu \nu} - (1 + \xi^{-1})
k^{\mu}k^{\nu} - \Pi^{\mu \nu}(k)]A_{\nu}^{a}(k)=
\]
\begin{equation}
= f^{abc} \int \, S_{k,k_{1},k_{2}}^{\mu \nu \lambda}
(A_{\nu}^{b}(k_{1})A_{\lambda}^{c}(k_{2}) -
\langle A_{\nu}^{b}(k_{1})A_{\lambda}^{c}(k_{2}) \rangle)
\delta(k - k_{1} - k_{2}) dk_{1}dk_{2}.
\label{eq:hh}
\end{equation}
The approximate solution of this equation is in the form
\[
A_{\mu}^{a}(k)=A_{\mu}^{(0)a}(k) - {\cal D}_{\mu \nu}(k)f^{abc} \int \,
S_{k,k_{1},k_{2}}^{\nu \lambda \sigma}(
A_{\lambda}^{(0)b}(k_{1})A_{\sigma}^{(0)c}(k_{2}) -
\]
\begin{equation}
- \langle A_{\lambda}^{(0)b}(k_{1})A_{\sigma}^{(0)c}(k_{2}) \rangle )
\delta(k - k_{1} - k_{2}) dk_{1}dk_{2},
\label{eq:jj}
\end{equation}
where $A_{\mu}^{(0)a}(k)$ is a solution of homogeneous Eq. (\ref{eq:hh})
corresponding noninteracting fields, and
\begin{equation}
{\cal D}_{\mu \nu}(k)=-[k^{2}g_{\mu \nu} - (1 + \xi^{-1})k_{\mu}k_{\nu} -
\Pi_{\mu \nu}(k)]^{-1}
\label{eq:kk}
\end{equation}
represents the medium modified (retarded) gluon propagator.

Now we substitute (\ref{eq:jj}) into correlators of three fields introducing
in Eq. (\ref{eq:gg}). Because $A^{(0)}$ represents amplitudes of entirely
uncorrelated waves, the correlation function with three $A^{(0)}$ drops out.
In this case it should be defined more exactly
each of the terms in $ \langle A_{\mu}^{\ast a}(k^{\prime})A_{\nu}^{c}
(k_{1})A_{\lambda}^{d}(k_{2}) \rangle $ and
$ \langle A_{\mu}^{b}(k)A_{\nu}^{\ast c}
(k_{1})A_{\lambda}^{\ast d}(k_{2}) \rangle $. In the correlation
functions of four amplitudes, within the accepted accuracy, it can be done
distinction between the fields $A$ and $A^{(0)}$.

Finally Eq. (\ref{eq:gg}) becomes
\[
\frac{\partial}{\partial k_{\lambda}}[k^{2}g^{\mu \nu} -
(1 + \xi ^{-1})k^{\mu}k^{\nu} - \Pi^{H \mu \nu}(k)]
\frac{\partial I_{\mu \nu}^{ab}}{\partial x^{\lambda}}=
\]
\[
= -i \int \, dk^{\prime} dk_{1} dk_{2} dk_{3} \{
f^{bcf}f^{fde} \delta(k - k_{1} - k_{2} - k_{3}) \tilde{\Sigma}_{k,
k_{1},k_{2},k_{3}}^{\mu \nu \lambda \sigma}( \langle
A_{\mu}^{\ast a}(k^{\prime})A_{\nu}^{c}(k_{3})A_{\lambda}^{d}(k_{1})
A_{\sigma}^{e}(k_{2}) \rangle -
\]
\[
- \langle A_{\mu}^{\ast a}(k^{\prime})A_{\nu}^{c}(k_{3}) \rangle
\langle A_{\lambda}^{d}(k_{1})A_{\sigma}^{e}(k_{2}) \rangle ) -
\]
\[
- f^{acf}f^{fde} \delta(k^{\prime} - k_{1} - k_{2} - k_{3})
\tilde{\Sigma}_{k^{\prime},
k_{1},k_{2},k_{3}}^{\ast \mu \nu \lambda \sigma}( \langle
A_{\mu}^{b}(k)A_{\nu}^{\ast c}(k_{3})A_{\lambda}^{\ast d}(k_{1})
A_{\sigma}^{\ast e}(k_{2}) \rangle -
\]
\begin{equation}
- \langle A_{\mu}^{b}(k)A_{\nu}^{\ast c}(k_{3}) \rangle
\langle A_{\lambda}^{\ast d}(k_{1})A_{\sigma}^{\ast e}(k_{2}) \rangle ) \} +
\label{eq:ll}
\end{equation}
\[
+ if^{bcd}f^{aef} \int \, dk^{\prime} \int \, dk_{1}dk_{2}
dk^{\prime}_{1}dk^{\prime}_{2} \, ({\cal D}^{\ast}_{\rho \alpha}(k^{\prime}) -
{\cal D}_{\alpha \rho}(k)) S_{k,k_{1},k_{2}}^{\rho \mu \nu}
S_{k^{\prime},k_{1}^{\prime},k_{2}^{\prime}}^{\ast \alpha \lambda \sigma} (
\langle A_{\mu}^{c}(k_{1})A_{\nu}^{d}(k_{2})
\]
\[
A_{\lambda}^{\ast e}(k_{1}^{\prime})A_{\sigma}^{\ast f}(k_{2}^{\prime})
\rangle - \langle A_{\mu}^{c}(k_{1})A_{\lambda}^{d}(k_{2}) \rangle
\langle A_{\lambda}^{\ast e}(k_{1}^{\prime})
A_{\sigma}^{\ast f}(k_{2}^{\prime}) \rangle )
\delta (k - k_{1} - k_{2}) \delta (k^{\prime} - k_{1}^{\prime}
- k_{2}^{\prime}).
\]
Here,
\begin{equation}
\tilde{\Sigma}_{k,k_{1},k_{2},k_{3}}^{\mu \nu \lambda \sigma}
\equiv \Sigma_{k,k_{1},k_{2},k_{3}}^{(I) \mu \nu \lambda \sigma} -
(S_{k,k_{3},k_{1} + k_{2}}^{\mu \nu \rho} -
S_{k,k_{1} + k_{2},k_{3}}^{\mu \rho \nu}){\cal D}_{\rho \alpha}(k_{1}
+k_{2})S_{k_{1} + k_{2},k_{1},k_{2}}^{\alpha \lambda \sigma}.
\label{eq:zz}
\end{equation}

It follows from the definition (\ref{eq:kk}) that the propagator is of order
$\sim 1/g^{2}T^{2}$. Taking into account (\ref{eq:ii}) and (\ref{eq:ff}) we see that
all terms in the right-hand side of (\ref{eq:ll}) are of the same order. This
explains why in the expansion of the current (\ref{eq:f}) the following term
- $j^{T(3)}$ should be retained in addition to the first nonlinear
correction $j^{T(2)}$: it leads to the effects of the same
quantity order.\\

{\bf 7. THE KINETIC EQUATION FOR LONGITUDINAL
WAVES}\\

Let us divide the mean of four fields in Eq. (\ref{eq:ll}) into three pairwise
products of two-point correlations by the rule
\[
\langle A(k_{1})A(k_{2})A(k_{3})A(k_{4}) \rangle =
\langle A(k_{1})A(k_{2}) \rangle \langle A(k_{3})A(k_{4}) \rangle +
\langle A(k_{1})A(k_{3}) \rangle \langle A(k_{2})A(k_{4}) \rangle +
\]
\[
+ \langle A(k_{1})A(k_{4}) \rangle \langle A(k_{2})A(k_{3}) \rangle.
\]

Taking into account that the spectral densities in the right-hand side
of Eq. (\ref{eq:ll}) can be considered as stationary and homogeneous those,
i.e. having the form (\ref{eq:b}), and setting $I_{\mu \nu}^{ab}= \delta^{ab} I_{\mu \nu}$, we find instead of
Eq. (\ref{eq:ll})
\[
\frac{\partial}{\partial k_{\lambda}}[k^{2}g^{\mu \nu} - (1 + \xi^{-1})
k^{\mu}k^{\nu} - \Pi^{H \mu \nu}(k)]
\frac{\partial I_{\mu \nu}}{\partial x^{\lambda}} =
\]
\begin{equation}
= 2N_{c} \int \, dk_{1} \, {\rm Im}(
\tilde{\Sigma}_{k,k,k_{1},- k_{1}}^{\mu \nu \lambda \sigma} -
\tilde{\Sigma}_{k,k_{1},k,- k_{1}}^{\mu \nu \sigma \lambda})
I_{\mu \lambda}(k) I_{\nu \sigma}(k_{1}) +
\label{eq:xx}
\end{equation}
\[
+ N_{c}{\rm Im}({\cal D}_{\rho \alpha}(k)) \int \, dk_{1}dk_{2}
\, (S_{k,k_{1},k_{2}}^{\rho \mu \nu} - S_{k,k_{2},k_{1}}^{\rho \nu \mu})
(S_{k,k_{1},k_{2}}^{\ast \alpha \lambda \sigma}
- S_{k,k_{2},k_{1}}^{\ast \alpha \sigma \lambda})I_{\mu \lambda}(k_{1})
I_{\nu \sigma}(k_{2}) \delta (k - k_{1} - k_{2}).
\]

As it is known [13, 14], in global equilibrium QGP the oscillations of three
typies can be extended: the longitudinal, transverse and
nonphysical oscillations. In this connection we define the spectral density
$I_{\mu \nu}(k,x)=I_{\mu \nu}$ in the form of expansion
\begin{equation}
I_{\mu \nu}=P_{\mu \nu}I_{k}^{t} + Q_{\mu \nu}I_{k}^{l} +
\xi D_{\mu \nu}I_{k}^{n} \; , \; I_{k}^{(t,l,n)} \equiv I^{(t,l,n)}(k,x),
\label{eq:cc}
\end{equation}
where the transverse and the longitudinal projectors [14] are $P_{\mu \nu}
= g_{\mu \nu} - k_{\mu}k_{\nu}/k^2 - Q_{\mu \nu} \, ,\, Q_{\mu \nu}=
\bar{u}_{\mu} \bar{u}_{\nu}/ \bar{u}^{2},$ and $\bar{u}_{\mu}=k^{2}u_{\mu}-
k_{\mu}(ku); D_{\mu \nu}=k_{\mu}k_{\nu}/k^2.$
By using these projectors the propagator (\ref{eq:kk}) can be written in
more convenient form
\begin{equation}
{\cal D}_{\mu \nu}(k)= -
\frac{P_{\mu \nu}}{k^{2} - \Pi^{t}} -
\frac{Q_{\mu \nu}}{k^{2} - \Pi^{l}} +
\xi \frac{D_{\mu \nu}}{k^{2} + i \epsilon}.
\label{eq:vv}
\end{equation}
Here, $\Pi^{t}= \frac{1}{2} \Pi^{\mu \nu}P_{\mu \nu}, \, \Pi^{l}=
\Pi^{\mu \nu}Q_{\mu \nu}$.
At finite temperature, the velocity of plasma introduces
a preferred direction in space-time which breaks manifest Lorentz invariance.
Let us assume that we are in the rest frame of the heat bath,
so that  $u_{\mu}=(1,0,0,0).$

Eq. (\ref{eq:xx}) with the expansions (\ref{eq:cc}) and (\ref{eq:vv}) enables us to
investigate various nonlinear processes in QGP:
the nonlinear scattering of longitudinal waves in longitudinal
or transverse waves; the scattering of transverse waves in longitudinal
or transverse waves; the merger of two longitudinal waves in one
transverse wave etc.. In this work we restrict our consideration to
investigation of most simple process - the nonlinear scattering
of longitudinal waves by particles of QGP in longitudinal those.

Further derivation of kinetic equation for longitudinal oscillations
is similar to corresponding derivation in the theory of electromagnetic
plasma, therefore we restrict our consideration to its schematic description.

Now we omit nonlinear terms and anti-Hermitian part of the polarization
tensor in Eq. (\ref{eq:c}). Further substituting the function
$\delta^{ab}Q_{\mu \nu}(k)I_{k}^{l} \delta(k^{\prime} - k)$ instead
of $I_{\mu \nu}^{ab}(k^{\prime},k)$, we lead to the equation
\[
{\rm Re} \, ( \varepsilon^{l}(k)) \, I_{k}^{l} =0.
\]
Here, we use relation: $1 - \Pi^{l}(k)/k^{2}= {\varepsilon}^{l}(k)$.
The solution of this equation has the structure
\begin{equation}
I_{k}^{l}=I_{\bf k}^{l} \delta ( \omega - {\omega}_{\bf k}^{l}) +
I_{- \bf k}^{l} \delta( \omega + {\omega}_{\bf k}^{l}) \; , \;
{\omega}_{\bf k}^{l} >0,
\label{eq:bb}
\end{equation}
where $I_{\bf k}^{l}$ is a certain function of a wave vector ${\bf k}$
and ${\omega}_{\bf k}^{l} \equiv {\omega}^{l}({\bf k})$ is a frequency
of the longitudinal eigenwaves in QGP.

The equation describing the variation of spectral density of
longitudinal oscillations is obtained from Eq. (\ref{eq:xx}) by replacement:
$I_{\mu \nu} \rightarrow Q_{\mu \nu}(k)I_{k}^{l}$, where $I_{k}^{l}$
is defined by (\ref{eq:bb}). $\delta$-functions in (\ref{eq:bb}) enable us to remove
integration over frequency and thus we have instead of Eq. (\ref{eq:xx})
\[
\left( k^{2} \frac{\partial {\rm Re} {\varepsilon}^{l}(k)}
{\partial k_{\lambda}} \right)_{\omega = \omega_{\bf k}^{l}}
\frac{\partial I_{\bf k}^{l}}{\partial x^{\lambda}}=
2N_{c}I_{\bf k}^{l} \int \, d{\bf k}_{1} \, I_{{\bf k}_{1}}^{l}
\big( {\rm Im} \,
[( \tilde{\Sigma}_{k,k,k_{1},-k_{1}}^{\mu \nu \lambda \sigma} -
\tilde{\Sigma}_{k,k_{1},k,-k_{1}}^{\mu \nu \sigma \lambda}) +
\]
\begin{equation}
+ ( \tilde{\Sigma}_{k,k,-k_{1},k_{1}}^{\mu \nu \lambda \sigma} -
\tilde{\Sigma}_{k,-k_{1},k,k_{1}}^{\mu \nu \sigma \lambda})]
Q_{\mu \lambda}(k)Q_{\nu \sigma}(k_{1}) \big)_{\omega= \omega_{\bf k}^{l}, \,
\omega_{1}= \omega_{{\bf k}_{1}}^{l}} +
\label{eq:nn}
\end{equation}
\[
+ N_{c} \int_{0}^{\infty} d \omega
\int \, d{\bf k}_{1} d{\bf k}_{2} I_{{\bf k}_{1}}^{l}
I_{{\bf k}_2}^{l} (
G_{k,k_1,k_2} + G_{k,-k_1,k_2} + G_{k,k_1,-k_2} + G_{k,-k_1,-k_2}
)_{\omega_{1}= \omega_{{\bf k}_{1}}^{l}, \,
\omega_{2}= \omega_{{\bf k}_{2}}^{l}} \, ,
\]
where
\[
G_{k,k_1,k_2} = {\rm Im}({\cal D}_{\rho \alpha}(k))
(S_{k,k_{1},k_{2}}^{\rho \mu \nu} - S_{k,k_{2},k_{1}}^{\rho \nu \mu})
(S_{k,k_{1},k_{2}}^{\ast \alpha \lambda \sigma}
- S_{k,k_{2},k_{1}}^{\ast \alpha \sigma \lambda})Q_{\mu \lambda}(k_{1})
Q_{\nu \sigma}(k_{2}) \delta(k - k_{1} -k_{2}).
\]

Let us consider the terms entering in the right-hand side of
Eq. (\ref{eq:nn}). The integral with the function $G_{k,k_{1},k_{2}}$ is different
from zero if the conservation laws are obeyed
\begin{equation}
{\bf k}={\bf k}_{1} + {\bf k}_{2},
\label{eq:mm}
\end{equation}
\[
\omega_{\bf k}^{l} = \omega_{{\bf k}_1}^{l} +
\omega_{{\bf k}_2}^{l}.
\]
These conservation laws describe a decay of one longitudinal wave in
two longitudinal waves. However for a spectrum of the longitudinal oscillations
in QGP, the equalities (\ref{eq:mm}) do not hold simultaneously, no matter what
the values of the wave vectors ${\bf k},{\bf k}_{1}$ and ${\bf k}_{2}$
may be, i.e. this nonlinear process is forbidded. Therefore the integral
with $G_{k,k_{1},k_{2}}$ vanishes. Remaining integrals with $G$-functions
differ from (\ref{eq:mm}) in that some of the interacting waves are not radiated
but absorped. They also vanish.

The expression
\begin{equation}
( \tilde{\Sigma}_{k,k,-k_{1},k_{1}}^{\mu \nu \lambda \sigma} -
\tilde{\Sigma}_{k,- k_{1},k,k_{1}}^{\mu \nu \sigma \lambda} ) \,
Q_{\mu \lambda}(k)Q_{\nu \sigma}(k_{1}) \mid_{\omega=
\omega_{\bf k}^{l} , \, \omega_{1}= \omega_{{\bf k}_1}^{l}},
\label{eq:qqq}
\end{equation}
contains the factors
\[
1/(pk + ip_{0} \epsilon) \, , \, 1/(pk_{1} + ip_{0} \epsilon) \, ,
\, 1/(p(k - k_{1}) + ip_{0} \epsilon),
\]
by the definitions of functions entering in it.
Imaginary parts of first two factors should be setting equal to
zero, because they are connected with linear Landau damping of
longitudinal waves (which is absent in QGP), and therefore the imaginary
part of the expression (\ref{eq:qqq}) properly introducing in Eq.
(\ref{eq:nn}) will be defined
as
\[
{\rm Im} \frac{1}{p(k - k_{1}) + ip_{0} \epsilon}
\bigg{\vert}_{\omega = \omega_{\bf k}^{l}, \,
\omega_{1}= \omega_{{\bf k}_{1}}^{l}} =
- \frac{i \pi}{p_{0}} \delta (\omega_{\bf k}^{l} - \omega_{{\bf k}_{1}}^{l}
- {\bf v}({\bf k} - {\bf k}_{1})).
\]
It follows that nonlinear term in the right-hand side of (\ref{eq:nn})
with the function
(\ref{eq:qqq}) is different from zero if the conservation law is obeyed
\[
\omega_{\bf k}^{l} - \omega_{{\bf k}_{1}}^{l}
- {\bf v}({\bf k} - {\bf k}_{1})=0.
\]
This conservation law describes the process of scattering of longitudinal
wave (plasmon) in longitudinal one by the particles in QGP.

Let us consider in more detail the term in (\ref{eq:qqq})
(see definition (\ref{eq:zz}))
with propagator ${\cal D}_{\rho \alpha}(k - k_{1})$. By expansion (\ref{eq:vv})
this propagator represents the nonlinear interaction of
longitudinal waves with longitudinal ones through three types of
intermediate oscillations: the transverse, longitudinal and nonphysical
oscillations depending on a gauge parameter. The term with
\[
\left( \frac{P_{\rho \alpha}(k_{2})}{k_{2}^{2} - \Pi^{t}(k_{2})} \right)_{
\omega = \omega_{\bf k}^{l}, \, \omega_{1} = \omega_{{\bf k}_{1}}^{l}}
\]
(hereafter $k_{2} \equiv k - k_{1}$) in general, describes two
fundamentally different nonlinear processes:
\begin{enumerate}
\item if $k_{2}=( \omega_{\bf k}^{l} - \omega_{{\bf k}_{1}}^{l},
{\bf k} - {\bf k}_{1})$ is a solution of the dispersion equation
$k_{2}^{2} - \Pi^{t}(k_{2})=0$, then this term describes the process of
merger of two longitudinal oscillations in transverse eigenwave;
\item otherwise, it defines the process of nonlinear scattering of
longitudinal waves in longitudinal those through the transverse virtual
oscillation (for a virtual wave in distinction to the eigenwave,
a frequency $\omega$ and a wave vector ${\bf k}$ are not connected with
each other by the dispersion dependence: $\omega \neq \omega({\bf k})$).
\end{enumerate}
The equality $k_{2}^{2} - \Pi^{l}(k_{2})=0$
does not hold for longitudinal oscillations,
as we see above and therefore, the term
\[
\left( \frac{Q_{\rho \alpha}(k_{2})}{k_{2}^{2} - \Pi^{l}(k_{2})} \right)_{
\omega = \omega_{\bf k}^{l}, \omega_{1} = \omega_{{\bf k}_{1}}^{l}}
\]
defines only the process of scattering of longitudinal waves in longitudinal
those through the longitudinal virtual oscillation.

Let us consider the contribution of nonphysical intermediate oscillations
\[
\left( \xi \, \frac{D_{\rho \alpha}(k_2)}{k_2^2 + i \epsilon }
\right)_{\omega= \omega_{\bf k}^{l}, \omega_1 = \omega_{{\bf k}_1}^{l}}.
\]
By direct calculation one can show, that in contraction with tensor
$D_{\rho \alpha}(k_2)$ the complex factor $1/(p(k - k_1) + i p_0 \epsilon$
is reduced in the expressions with $S$-functions. In particular it follows
that contribution from the process of nonlinear scattering
of longitudinal waves,
connected with nonphysical intermediate oscillations drops out.
Therefore the gauge parametor $\xi$ is absent in the equation for
longitudinal wave.

The remaining terms with $\tilde{\Sigma}$ are distinguished from above
considered terms by a sign of $k_{1}$, and describe the processes of simultaneous
radiation or absorption by particles of two waves. The contribution
of these processes is exponentially small in relation to the scattering process
and therefore these terms are omitted.

Summing the preceding and going from the function $I_{\bf k}^{l}$
to the function
\[
W_{\bf k}^{l}= - \left( \omega k^{2} \frac{\partial {\rm Re}
\varepsilon^{l}(k)}{\partial \omega} \right)_{\omega =
\omega_{\bf k}^{l}} I_{{\bf k}}^{l} \, ,
\]
having the physical meaning of spectral density of energy of
longitudinal oscillations, we find from (\ref{eq:nn}) the required kinetic
equation for longitudinal waves in QGP
\begin{equation}
\frac{\partial W_{\bf k}^{l}}{\partial t} +
{\bf V}_{\bf k}^{l} \, \frac{\partial W_{\bf k}^{l}}
{\partial {\bf x}} = - \hat{\gamma} \{
( \frac{W_{\bf k}^{l}}{\omega_{\bf k}^{l}})  \} \, W_{\bf k}^{l},
\label{eq:www}
\end{equation}
where
\[
{\bf V}_{\bf k}^{l} = \frac{\partial \omega_{\bf k}^{l}}
{\partial {\bf k}} =
- \Big[ \big( \frac{\partial {\rm Re} \,
\varepsilon^{l}(k)}{\partial {\bf k}} \big)
\Big/ \big( \frac{\partial {\rm Re} \, \varepsilon^{l}(k)}
{\partial \omega} \big) \Big]_{\omega = \omega_{\bf k}^{l}}
\]
is the group velocity of longitudinal oscillations and
\[
\hat{\gamma} \{ \Big( \frac{W_{\bf k}^{l}}{\omega_{\bf k}^{l}} \Big) \}
\equiv \gamma^{l}({\bf k}) = 2N_{c} \int \, d{\bf k}_{1}
\Big( \frac{W_{{\bf k}_{1}}^{l}}{\omega_{{\bf k}_{1}}^{l}} \Big)
\bigg[ \frac{1}{k^{2}k^{2}_{1}}
\Big( \frac{\partial {\rm Re} \, \varepsilon^{l}(k)}
{\partial \omega} \Big)^{-1}
\Big( \frac{\partial {\rm Re} \, \varepsilon^{l}(k_{1})}
{\partial \omega_{1}} \Big)^{-1}
\]
\begin{equation}
{\rm Im}( \tilde{\Sigma}_{k,k,-k_{1},k_{1}}^{\mu \nu \lambda \sigma} -
\tilde{\Sigma}_{k,-k_{1},k,k_{1}}^{\mu \nu \sigma \lambda})
Q_{\mu \lambda}(k)Q_{\nu \sigma}(k_{1}) \bigg]_{\omega = \omega_{\bf k}^{l},
\, \omega_{1}= \omega_{{\bf k}_{1}}^{l}}
\label{eq:eee}
\end{equation}
presents the damping rate caused by nonlinear effects and being the linear
functional of spectral density of energy.

One can write (\ref{eq:www}) in the form which is more close to usual representation
if the spectral density of number of longitudinal oscillations is
entered
\[
N_{\bf k}^{l} = W_{\bf k}^{l}/ \omega_{\bf k}^{l}.
\]
It fulfils the role of distribution function of a number of plasmons.
Then instead of (\ref{eq:www}) we have
\begin{equation}
\frac{{\rm d} N_{\bf k}^{l}}{{\rm d}t} \equiv
\frac{\partial N_{\bf k}^{l}}{\partial t} +
{\bf V}_{\bf k}^{l} \, \frac{\partial N_{\bf k}^{l}}
{\partial {\bf x}} = - \hat{\gamma} \, \{
N_{\bf k}^{l} \} \, N_{\bf k}^{l} \, .
\label{eq:rrr}
\end{equation}

{\bf 8. THE PHYSICAL MECHANISM OF THE NONLINEAR
SCATTERING OF WAVES}\\

Now we transform $\gamma^{l}({\bf k})$ to the form allowing more neatly
explain the physical meaning of the terms entering in the nonlinear
damping rate. The first transformation of this type was proposed by Tsytovich
for electromagnetic plasma [11].

By the definition $\tilde{\Sigma}$ (\ref{eq:zz}) we have
\[
( \tilde{\Sigma}_{k,k,-k_{1},k_{1}}^{\mu \nu \lambda \sigma}
- \tilde{\Sigma}_{k,-k_{1},k,k_{1}}^{\mu \nu \sigma \lambda})
Q_{\mu \lambda}(k)Q_{\nu \sigma}(k_{1}) =
(\Sigma_{k,k,-k_{1},k_{1}}^{(I) \mu \nu \lambda \sigma}
- \Sigma_{k,-k_{1},k,k_{1}}^{(I) \mu \nu \sigma \lambda})
Q_{\mu \lambda}(k)Q_{\nu \sigma}(k_{1}) +
\]
\begin{equation}
+ \frac{1}{k_{2}^{2} - \Pi^{l}(k_{2})} \,
(S_{k,k_{1},k_{2}}^{\mu \nu \rho} - S_{k,k_{2},k_{1}}^{\mu \rho \nu})
(S_{k_{2},k,-k_{1}}^{\alpha \lambda \sigma} -
S_{k_{2},-k_{1},k}^{\alpha \sigma \lambda})Q_{\rho \alpha}(k_{2})
Q_{\mu \lambda}(k)Q_{\nu \sigma}(k_{1}).
\label{eq:ttt}
\end{equation}
Here, for simplicity in the last term in the right-hand side of Eq.
(\ref{eq:ttt})
the effects connected with existence of transverse intermediate
oscillations are neglected. At first we consider the expression with
$\Sigma^{(I)}$. By virtue of definition (\ref{eq:ss}) we have
\[
{\rm Im} \, ( \Sigma_{k,k,-k_{1},k_{1}}^{(I) \mu \nu \lambda \sigma}
- \Sigma_{k,-k_{1},k,k_{1}}^{(I) \mu \nu \sigma \lambda})
Q_{\mu \lambda}(k)Q_{\nu \sigma}(k_{1}) =
\]
\[
= - \frac{\pi g^{4}}{{\bf k}^{2}{\bf k}_{1}^{2}k^{2}k_{1}^{2}}
\int \, d^{4}p \, \frac{(p \bar{u}(k))^{2}(p \bar{u}(k_{1}))^{2}}
{(pk)^{2}} \, \delta (pk_{2}) (k_{2} \partial_{p} {\cal N}_{eq}).
\]

The contribution to $\gamma^{l}({\bf k})$ from this term may be
introduce in the form
\begin{equation}
- \int \, {\it w}_{p}^{\Sigma}({\bf k}, {\bf k}_{1}) N_{{\bf k}_1}^{l}
p_{0} (k_{2} \partial_{p} {\cal N}_{eq}) \, d^{4}pd{\bf k}_{1},
\label{eq:yyy}
\end{equation}
where
\[
{\it w}_{p}^{\Sigma}({\bf k}, {\bf k}_{1}) =
\]
\begin{equation}
= \frac{2 \pi N_{c}}{(k^{2}k^{2}_{1})^{2}}
\bigg( \frac{\partial {\rm Re} \, \varepsilon^{l}(k)}{\partial \omega}
\bigg)_{\omega = \omega_{\bf k}^{l}}^{-1}
\bigg( \frac{\partial {\rm Re} \, \varepsilon^{l}(k_{1})}{\partial \omega_{1}}
\bigg)_{\omega_{1} = \omega_{\bf k_1}^{l}}^{-1}
\delta ( \omega_{\bf k}^{l} - \omega_{{\bf k}_{1}}^{l} -
{\bf v}( {\bf k} - {\bf k}_{1}))
\vert \Lambda^{\Sigma}({\bf k},{\bf k}_{1}) \vert^{2},
\label{eq:uuu}
\end{equation}
\begin{equation}
\Lambda^{\Sigma}({\bf k},{\bf k}_{1}) =
\frac{g^{2}}{ \vert {\bf k} \vert \vert {\bf k}_{1} \vert}
\, \frac{[ \omega_{\bf k}^{l}({\bf k}{\bf v}) - {\bf k}^{2}]
[ \omega_{{\bf k}_{1}}^{l}({{\bf k}_{1}}{\bf v}) - {\bf k}^{2}_{1}]}
{\omega_{\bf k}^{l} - ({\bf k}{\bf v})}.
\label{eq:iii}
\end{equation}

To clear up the physical meaning of contribution (\ref{eq:yyy}), it is convinient
to compare it with appropriate contribution
in the theory of electromagnetic plasma.
In this case as shown in [11] this contribution describes the Thomson scattering
of a wave $\omega_{\bf k}^{l}$ by particles: a wave $\omega_{\bf k}
^{l}$ sets particles of plasma into oscillation and oscillating particles
radiate a wave $\omega_{{\bf k}_{1}}^{l}$. The corresponding function
${\it w}_{p}^{\Sigma}({\bf k},{\bf k}_{1})$ presents the probability of
this scattering. As it was shown above in quark-gluon plasma for a soft
long-wavelength excitations all Abelian contributions is at most $g \ln g$
times the non-Abelian ones and the basic scattering mechanism here, is
essentially another (next we consider it briefly, the details will be
published elsewhere).

For revealing this mechanism we use the classical pattern of QGP
description [2], in which the particles states are characterized besides
coordinate and momentum by the color vector ${\rm Q}= (
{\rm Q}^{a}), a=1, \ldots,
N_{c}^{2} - 1 $ also. As was shown by Heinz [2] there is an intimate
connection between the classical kinetic equations and semiclassical
ones (\ref{eq:r}). Therefore in this case use of classical notions is
justified.

Let the field acting on a color particle in QGP represents
a set of longitudinal plane waves
\begin{equation}
\tilde{A}_{\mu}^{a}(x) = \int \, [Q_{\mu \nu}(k)A_{\bf k}^{a \nu}]_{\omega=
\omega_{\bf k}^{l}} {\rm e}^{i {\bf k}{\bf x} - i \omega_{\bf k}^{l}t}
d {\bf k}.
\label{eq:ooo}
\end{equation}
The particle motion in this wave field is described by the system of
equations
\begin{equation}
{\rm m} \frac{{\rm d}^{2}x^{\mu}}{{\rm d} \tau^{2}} =
g {\rm Q}^{a} \tilde{F}^{a \mu \nu} \frac{{\rm d}x_{\nu}}{{\rm d} \tau},
\label{eq:ppp}
\end{equation}
\begin{equation}
\frac{{\rm d}{\rm Q}^{a}}{{\rm d} \tau}=
- gf^{abc} \frac{{\rm d}x^{\mu}}{{\rm d} \tau} \tilde{A}_{\mu}^{b}
{\rm Q}^{c}.
\label{eq:aaa}
\end{equation}
Here, $\tau$ is a proper time of a particle.
The system (\ref{eq:ppp}), (\ref{eq:aaa}) is solved
by the method of successive approximations - expansion in the field
amplitude. A zeroth approximation describes uniform restlinear motion,
and the next one - constrained charge oscillations in the field
(\ref{eq:ooo}). With a knowledge of the motion law of a charge, the intensity
of radiating by it longitudinal waves can be defined. In this case Eq.
(\ref{eq:ppp})
defines the Abelian contribution to radiation, whereas (\ref{eq:aaa}) - non-Abelian
one and interference of these two contributions equals zero. The scattering
probability calculated by this means, based on Eq. (\ref{eq:aaa}) is coincident with
obtained above (\ref{eq:uuu}).

In this manner the contribution (\ref{eq:yyy}) to $\gamma^{l}({\bf k})$ are caused
 by not the spatially oscillations of a color particle, as it occurs in
electromagnetic plasma, but the initiation of a precession of a color
vector ${\rm Q}$ of a particle in field of a longitudinal wave (\ref{eq:ooo})
(Eq. (\ref{eq:aaa}) conserves the length of a color vector).

Let us consider now more complicated term in (\ref{eq:ttt}) connected with
$S$-functions. By exact calculation, using the definitions (\ref{eq:x}) and
(\ref{eq:tt}), it is not difficult to see that the following equality is obeyed
\[
(S_{k,k_{1},k_{2}}^{\mu \nu \rho} -
S_{k,k_{2},k_{1}}^{\mu \rho \nu }) \, \bar{u}_{\mu}(k) \bar{u}_{\rho}(k_2)
\bar{u}_{\nu}(k_{1}) =
- \, (S_{k_{2},k,-k_{1}}^{\alpha \lambda \sigma} -
S_{k_{2},-k_{1},k}^{\alpha \sigma \lambda}) \,
\bar{u}_{\alpha}(k_{2}) \bar{u}_{\lambda}(k)
\bar{u}_{\sigma}(k_{1}) \equiv
\]
\begin{equation}
\equiv S_{k,k_{1}} = S_{k,k_{1}}^{(I)} + S_{k,k_{1}}^{(II)}.
\label{eq:sss}
\end{equation}
Then the contribution to $\gamma^{l}({\bf k})$ from $S$-functions
can be represented as
\[
2N_{c} \int \,d{{\bf k}_{1}} N_{{\bf k}_{1}}^{l}
\frac{1}{{\bf k}^{2}{\bf k}_{1}^{2}{\bf k}_{2}^{2}}
\Big( \frac{\partial {\rm Re} \, \varepsilon^{l}(k)}{\partial \omega}
\Big)_{\omega = \omega_{\bf k}^{l}}^{-1}
\Big( \frac{\partial {\rm Re} \, \varepsilon^{l}(k_{1})}{\partial \omega_1}
\Big)_{\omega_{1} = \omega_{{\bf k}_{1}}^{l}}^{-1}
\]
\[
\Big[ \frac{1}{(k^{2}k_{1}^{2}k_{2}^{2})^2} {\rm Im} \,
\Big( \frac{1}{\varepsilon^{l}(k_{2})}(S_{k,k_{1}})^{2} \Big)
\Big]_{\omega = \omega_{\bf k}^{l}, \, \omega_{1} = \omega_{{\bf k}_{1}}^{l}}.
\]
Next we use the relation
\begin{equation}
{\rm Im}  \Big( \frac{1}{\varepsilon^{l}(k_{2})}(S_{k,k_{1}})^{2} \Big) =
\frac{{\rm Im} \,  \varepsilon^{l}(k_{2})}{\vert \varepsilon^{l}(k_{2})
\vert^{2}} \vert S_{k,k_{1}} \vert^{2} -
2 \, {\rm Im} (-iS_{k,k_{1}}) \,{\rm Re} \Big(
\frac{-iS_{k,k_{1}}}{ \varepsilon^{l}(k_{2})} \Big).
\label{eq:ddd}
\end{equation}
Taking into consideration the equality
\[
{\rm Im} \, \varepsilon^{l}(k_{2}) =
- \frac{\pi g^{2}}{{\bf k}_{2}^{2}} \int \, d^{4}p \, (p_{0})^{2}
\delta(pk_{2}) (k_{2} \partial_{p} {\cal N}_{eq}) \, ,
\]
one can write contribution from the first term in the right-hand side
of (\ref{eq:ddd}) to the nonlinear damping rate in the form
\begin{equation}
- \int \, {\it w}_{p}^{S}({\bf k}, {\bf k}_{1}) N_{{\bf k}_1}^{l}
p_{0} (k_{2} \partial_{p} {\cal N}_{eq}) \, d^{4}p \, d{\bf k}_{1},
\label{eq:fff}
\end{equation}
where
\[
{\it w}_{p}^{S}({\bf k}, {\bf k}_{1}) =
\]
\begin{equation}
= \frac{2 \pi N_{c}}{(k^{2}k^{2}_{1})^{2}}
\bigg( \frac{\partial {\rm Re} \, \varepsilon^{l}(k)}{\partial \omega}
\bigg)_{\omega = \omega_{\bf k}^{l}}^{-1}
\bigg( \frac{\partial {\rm Re} \, \varepsilon^{l}(k_{1})}{\partial \omega_{1}}
\bigg)_{\omega_{1} = \omega_{\bf k_1}^{l}}^{-1}
\delta ( \omega_{\bf k}^{l} - \omega_{{\bf k}_{1}}^{l} -
{\bf v}( {\bf k} - {\bf k}_{1}))
\vert \Lambda^{S}({\bf k},{\bf k}_{1}) \vert^{2},
\label{eq:ggg}
\end{equation}
\[
\Lambda^{S}({\bf k},{\bf k}_{1}) \equiv
\Lambda^{S^(I)}({\bf k},{\bf k}_{1}) + \Lambda^{S^(II)}({\bf k},{\bf k}_{1}) =
\]
\begin{equation}
= \frac{g}{\vert {\bf k} \vert \, \vert {\bf k}_{1} \vert \, {\bf k}_{2}^{2}}
\left( \frac{1}{k_{2}^{2}} \, \frac{-i (S_{k,k_{1}}^{(I)} + S_{k,k_{1}}^{(II)})}
{\varepsilon^{l}(k_{2})}
\right)_{\omega = \omega_{\bf k}^{l}, \, \omega_1 = \omega_{{\bf k}_{1}}^{l}}.
\label{eq:hhh}
\end{equation}

Here, $\Lambda^{S}$ represents the sum of two
contributions. The first contribution defining by the function $S_{k,k_{1}}^{(I)}$
is connected with self-action of a gauge field and had no the analogy in
electromagnetic plasma. The second contribution determining by the function
$S_{k,k_{1}}^{(II)}$ is associated with the scattering of wave by the
Debye screening shell of the particle. However, in contrast to
the electromagnetic plasma here, the scattering is accounted for not through
the oscillation of a coherent polarization cloud which surrounds the quark
as a result of interaction with
impingind wave, but as a consequence of the initiation of precession of
color vectors with particles forming this cloud in the wave field
$\omega_{\bf k}^{l}$. This process of scattering represents a pure
collective effect. For calculation of its probability it is necessary to
solve the kinetic equation describing a charge motion in a polarization
cloud in the field which is equal to the sum of fields of impinding wave
(\ref{eq:ooo}) and a charge producing a screening cloud.

The remaining term in the right-hand side of (\ref{eq:ddd}) describes the interference
of the above-mentioned scattering mechanisms. It is easily to see this having
used
\[
{\rm Im} \, (-iS_{k,k_{1}}) =
\pi g^{3} k_{2}^{2} \int \, d^{4}p \, \frac{p_{0}}{pk} \,
(p \bar{u}(k))(p \bar{u}(k_{1})) \delta(pk_{2})(k_{2} \partial_{p}
{\cal N}_{eq}).
\]
Thus, summing the preceding, instead of (\ref{eq:eee}) we have
\begin{equation}
\gamma^{l}({\bf k}) = - \int \, ( \omega_{\bf k}^{l} - \omega_{{\bf k}_
{1}}^{l}) Q_{p}({\bf k},{\bf k}_{1}) \Big( \frac{
W_{{\bf k}_{1}}^{l}}{\omega_{{\bf k}_1}^{l}} \Big) p_{0}
\frac{{\rm d}{\cal N}_{eq}(p_0)}{{\rm d}p_{0}} \, d^{4}p \, d{\bf k}_{1},
\label{eq:jjj}
\end{equation}
where
\begin{equation}
Q_{p}({\bf k}, {\bf k}_{1}) =
\frac{2 \pi N_{c}}
{[( \omega_{\bf k}^{l})^{2} - {\bf k}^{2}]^{2}
[( \omega_{{\bf k}_{1}}^{l})^{2} - {{\bf k}_{1}}^{2}]^{2}}
\bigg( \frac{\partial {\rm Re} \, \varepsilon^{l}(k)}{\partial \omega}
\bigg)_{\omega = \omega_{\bf k}^{l}}^{-1}
\bigg( \frac{\partial {\rm Re} \, \varepsilon^{l}(k_{1})}{\partial \omega_{1}}
\bigg)_{\omega_{1} = \omega_{\bf k_1}^{l}}^{-1}
\label{eq:kkk}
\end{equation}
\[
\delta ( \omega_{\bf k}^{l} - \omega_{{\bf k}_{1}}^l -
{\bf v}( {\bf k} - {\bf k}_{1}))
\vert \Lambda^{\Sigma}({\bf k},{\bf k}_{1})
+ \Lambda^{S}({\bf k},{\bf k}_{1}) \vert^{2},
\]
and the expressions for $\Lambda^{\Sigma}$ and $\Lambda^{S}$ are given by
(\ref{eq:iii})
and (\ref{eq:hhh}), respectively.

Now we note that it is convenient to interpret the terms entering in the
$\Lambda \equiv \Lambda^{\Sigma} + \Lambda^{S^{(I)}}
+ \Lambda^{S^{(II)}}$ by using a quantum language. In this case the term
$\Lambda^{\Sigma}$ connected with the Thomson scattering can be represented
as the Compton scattering of the oscillation quantum (plasmon) by QGP particle.
$\Lambda^{S(I)}$ defines the scattering of a
plasmon through a virtual wave with the propagator $1/(k_{2}^{2} -
\Pi^{l}(k_{2}))$, where a vertex of a three-wave interaction is induced by
a self-action of a field. The term $\Lambda^{S(II)}$ defines
the plasmon scattering by a particle through a virtual wave with the same
propagator and effective vertex of three-wave interaction connected
with the medium effects. In this case $\Lambda$ fulfils role
of the scattering amplitude.

The kernel $Q_{p}({\bf k},{\bf k}_{1})$ possesses two main properties. The
following inequality results from definition (\ref{eq:kkk})
\begin{equation}
Q_{p}({\bf k},{\bf k}_{1}) \geq 0 \, .
\label{eq:lll}
\end{equation}
Next, from (\ref{eq:iii}) it follows that $\Lambda^{\Sigma}({\bf k},{\bf k}_{1})=
\Lambda^{\ast \Sigma}({\bf k}_{1},{\bf k})$. The correctness of this equality
results from the conservation law: $\omega_{\bf k}^{l} - \omega_{{\bf k}_{1}}^
{l} - {\bf v}({\bf k} - {\bf k}_{1}) = 0$. For $\Lambda^{S}$ we have the
similar relation: $\Lambda^{S}({\bf k},{\bf k}_{1})=
\Lambda^{\ast S}({\bf k}_{1},{\bf k})$. Its proof trivially follows from the
definitions of $S_{k,k_{1}}^{(I)}, \, S_{k,k_{1}}^{(II)}$ and
$\varepsilon^{l}(k)$:
\begin{equation}
-iS_{k,k_{1}}^{(I)} = g \{
((k + k_{1}) \bar{u}(k_{2}))( \bar{u}(k) \bar{u}(k_{1})) -
(k_{1} \bar{u}(k))( \bar{u}(k_{1}) \bar{u}(k_{2})) -
2(k \bar{u}(k_{1}))( \bar{u}(k) \bar{u}(k_{2})) \},
\label{eq:zzz}
\end{equation}
\begin{equation}
- iS_{k,k_{1}}^{(II)} = -g^{3} \int \, d^{4}p^{\prime} \,
\frac{(p^{\prime} \bar{u}(k_{2}))(p^{\prime} \bar{u}(k))
(p^{\prime} \bar{u}(k_{1}))}
{p^{\prime}k_{2} + ip_{0}^{\prime} \epsilon}
\bigg(
\frac{(k \partial_{p^{\prime}}{\cal N}_{eq})}
{p^{\prime}k} -
\frac{(k_{1} \partial_{p^{\prime}}{\cal N}_{eq})}
{p^{\prime}k_{1}} \bigg),
\label{eq:xxx}
\end{equation}
\begin{equation}
\varepsilon^{l}(k) = 1 + \frac{3 \omega_{pl}^{2}}{{\bf k}^{2}}
\big[ 1 - F( \frac{\omega}{\vert {\bf k} \vert})  \big] \; , \;
F(x) = \frac{x}{2} \big[ \ln \bigg{\vert} \frac{1 + x}{1 - x} \bigg{\vert} -
i \pi \theta(1 - \vert x \vert) \big].
\label{eq:ccc}
\end{equation}

The consequense of these equalities is a main property of a symmetry kernel
$Q_{p}({\bf k},{\bf k}_{1})$ with respect to
permutation of a wave vectors ${\bf k}$
and ${\bf k}_{1}$
\begin{equation}
Q_{p}({\bf k},{\bf k}_{1}) =
Q_{p}({\bf k}_{1},{\bf k}).
\label{eq:vvv}
\end{equation}

Now let us consider consequence of the properties (\ref{eq:lll})
and (\ref{eq:vvv}).
Integrating (\ref{eq:rrr}) over $d{\bf k}/(2 \pi)^{3}$,
and using (\ref{eq:vvv}), we find
\[
\frac{{\rm d} N^{l}}{{\rm d}t} \equiv \frac{{\rm d}}{{\rm d}t}
\Big( \int \, N_{\bf k}^{l} \, \frac{d {\bf k}}{(2 \pi)^{3}} \Big)=0,
\]
i.e. the general number of plasmons $N^{l}$ in the process of the nonlinear
scattering conserves exactly.

From (\ref{eq:jjj}) it follows that in the case of a global equilibrium plasma,
when
\[
\frac{{\rm d}{\cal N}_{eq}(p_{0})}{{\bf d}p_{0}} < 0 \, ,
\]
and by inequality (\ref{eq:lll}),
waves of a high frequencies are damped out, and a smaller ones are increased.
Thus, due to the nonlinear interaction of a plasma oscillations a
pumping-over from short to long waves occurs in the spectrum, practically
leaving its total energy fixed. Therefore, the nonlinear decrement
$\gamma^{l}({\bf k})$ defines actually the inverse time of a spectral
pumping.

In the limit of $\vert {\bf k} \vert \rightarrow 0$ from (\ref{eq:jjj}) we obtain
\begin{equation}
\gamma^{l}(0) = - \int \, ( \omega_{pl} - \omega_{{\bf k}_
{1}}^{l}) Q_{p}(0,{\bf k}_{1})
\Big( \frac{W_{{\bf k}_{1}}^{l}}{\omega_{{\bf k}_{1}}^{l}} \Big) p_{0}
\frac{{\rm d}{\cal N}_{eq}(p_{0})}{{\rm d}p_{0}} \, d^{4}p \, d{\bf k}_{1}.
\label{eq:bbb}
\end{equation}

The frequency $\omega_{{\bf k}_{1}}^{l}$ is monotonically increasing function
of a wave number, therefore $\omega_{pl} - \omega_{{\bf k}_{1}}^{l} \leq 0$.
From this following inequality folows
\[
\gamma^{l}(0) <0,
\]
i.e. $\vert {\bf k} \vert =0$-mode is not damped, in contrast to
Ref. [8]. From the physical point of view this result is clear.
As it was shown above, nonlinear interaction of waves leads to the
spectral pumping from short to long waves. The mode $\vert {\bf k} \vert
=0 $ is a limiting in this series, and therefore $\gamma^{l}(0) < 0$.
It is clear that in the region $\vert {\bf k} \vert \simeq 0$ the other
nonlinear mechanisms switching off instability are to come
into effect. One of such mechanism, connected with regard to the
terms of a higher-degree of nonlinearity in expansion of the color
current (\ref{eq:f}) will be discussed in Conclusion.\\

{\bf 9. THE ESTIMATE OF $\gamma^{l}(0)$}\\

Now we consider the quantity $\gamma^{l}(0)$. As in the paper [8] it is
convenient to introduce the cylindrical coordinate system in which the
direction of polar axis is selected to be the same as the direction
${\bf k}_{1}$. Then the coordinates for ${\bf k},{\bf p},$ and
${\bf p}^{\prime}$ are ${\bf k} = ( \vert {\bf k} \vert, \alpha, \beta), \,
{\bf p} = ( \vert {\bf p} \vert, \theta, \varphi ),$
and ${\bf p}^{\prime} = ( \vert {\bf p}^{
\prime} \vert , \theta^{\prime}, \varphi^{\prime})$ correspondely. By $\phi$ and
$\phi^{\prime}$ we denote the corresponding angles between ${\bf p}$ and
${\bf k}$, ${\bf p}^{\prime}$ and ${\bf k}$. They can be expressed as
\begin{equation}
\cos \phi = \sin \theta \sin \alpha \cos( \varphi - \beta) + \cos \theta
\cos \alpha.
\label{eq:nnn}
\end{equation}
The expression for $\cos \phi^{\prime}$ is similar to above equation.

In the limit of $\vert {\bf k} \vert \rightarrow 0$ the kernel (\ref{eq:kkk})
transforms
to
\begin{equation}
Q_{p}(0,{\bf k}_1) =
\frac{\pi N_{c}}{\omega^{3}_{pl} \vert {\bf k}_1 \vert}
\, \frac{1}{[( \omega_{{\bf k}_{1}}^{l})^2 - {\bf k}_{1}^{2}]^{2}}
\Big( \frac{\partial {\rm Re} \, \varepsilon^{l}(k_{1})}{\partial \omega_{1}}
\Big)^{-1}_{\omega_{1}= \omega_{{\bf k}_{1}}^{l}}
\delta( \cos \theta - \rho_{{\bf k}_{1}}^{l})
\vert \Lambda^{\Sigma}(0,{\bf k}_1) +
\Lambda^{S}(0,{\bf k}_1) \vert^{2},
\label{eq:mmm}
\end{equation}
where
\[
\rho_{{\bf k}_1}^{l} \equiv ( \omega_{{\bf k}_1}^{l} - \omega_{pl})/
\vert {\bf k}_1 \vert \geq 0.
\]
Using definitions (\ref{eq:iii}) and (\ref{eq:hhh}), we find the expressions $\Lambda
^{\Sigma}(0,{\bf k}_1)$ and $\Lambda^{S}(0,{\bf k}_1)$.

Turning $\vert {\bf k} \vert$ to zero in (\ref{eq:iii}), we
have
\begin{equation}
\Lambda^{\Sigma}(0, {\bf k}_1) = g^2 \vert {\bf k}_1 \vert
\cos \phi \, (v_{{\bf k}_1}^{l} \cos \theta - 1).
\label{eq:qqqq}
\end{equation}
Here, we denote the phase velocity of longitudinal oscillations by
$v_{{\bf k}_1}^{l} = \omega_{{\bf k}_1}^{l}/ \vert {\bf k}_1 \vert$.

Calculation of $\Lambda^{S}(0,{\bf k}_{1})$ is more complicated.
From (\ref{eq:hhh})
we obtain
\begin{equation}
\Lambda^{S}(0,{\bf k}_{1}) = - \,
\frac{g}{\vert {\bf k}_1 \vert^3} \, \frac{1}{1 - \rho_{{\bf k}_1}^{l}} \,
\frac{1}{{\bf k}_1^2 + 3 \omega_{pl}^2 [1 - F(- \rho_{{\bf k}_1}
^{l})]}
\label{eq:wwww}
\end{equation}
\[
\lim_{\vert {\bf k} \vert \rightarrow 0}
\frac{-i}{\vert {\bf k} \vert}(S_{k,k_1}^{(I)} + S_{k,k_1}^{(II)})_{\omega =
\omega_{\bf k}^{l}, \, \omega_1 = \omega_{{\bf k}_1}^{l}}.
\]
Here, we use definition of the function $\varepsilon^l(k)$ (\ref{eq:ccc}).
Dividing (\ref{eq:zzz}) by
$\vert {\bf k} \vert$ and going to the limit $\vert {\bf k} \vert
\rightarrow 0$, after the simple, but slightly cumbersome computations,
we define
\begin{equation}
\lim_{\vert {\bf k} \vert \rightarrow 0}
\frac{-i S_{k,k_1}^{(I)}}{\vert {\bf k} \vert} \bigg{\vert}_{\omega =
\omega_{\bf k}^{l}, \, \omega_1 = \omega_{{\bf k}_1}^{l}} =
2g \cos \alpha \, \omega_{pl} \vert {\bf k}_1 \vert^3
\{ \omega_{pl}^2 - {\bf k}_1^2 (1 - v_{{\bf k}_1}^{l} \rho_{{\bf k}_1}^{l}) \}.
\label{eq:eeee}
\end{equation}

For definition of the limit of $(-iS_{k,k_1}^{(II)})/ \vert {\bf k} \vert$,
instead of (\ref{eq:xxx}) we use expression, which is defined from it, if the following
identity is accounted for
\[
\frac{1}{p^{\prime}k_{2} + ip_{0}^{\prime} \epsilon} \,
\frac{1}{p^{\prime} k_1} =
\Big( \frac{1}{p^{\prime}k_{2} + ip_{0}^{\prime} \epsilon} +
\frac{1}{p^{\prime}k_{1}} \Big) \, \frac{1}{p^{\prime}k}.
\]
This replacement leads to more simple limiting expression
\[
\lim_{\vert {\bf k} \vert \rightarrow 0}
\frac{-i S_{k,k_1}^{(II)}}{\vert {\bf k} \vert} \bigg{\vert}_{\omega =
\omega_{\bf k}^{l}, \, \omega_1 = \omega_{{\bf k}_1}^{l}} =
\frac{3g}{4 \pi} \omega_{pl}^2 \int_{0}^{2 \pi}  d \varphi^{\prime}
\int_{-1}^{1} \, d( \cos \theta^{\prime}) \cos \phi^{\prime}
{\bf k}_{1}^{4} (v_{{\bf k}_1}^{l} \cos \theta^{\prime} - 1)
( \rho_{{\bf k}_1}^{l} \cos \theta^{\prime} - 1)
\]
\begin{equation}
\left( \frac{\rho_{{\bf k}_1}^{l}}{\rho_{{\bf k}_1}^{l} - \cos \theta^{\prime}
- i \epsilon} - \frac{v_{{\bf k}_1}^{l}}{v_{{\bf k}_1}^{l} -
\cos \theta^{\prime}} \right).
\label{eq:rrrr}
\end{equation}
Here, we take into account that in view of definition of the equilibrium
function ${\cal N}_{eq}$ and (\ref{eq:j}) (for $\mu = 0$):
\[
\int_{- \infty}^{+ \infty} \, \vert {\bf p}^{\prime} \vert^{2} d \vert
{\bf p}^{\prime} \vert
\int_{- \infty}^{+ \infty}  p_{0}^{\prime} dp_{0}^{\prime} \,
\frac{{\rm d}{\cal N}_{eq}(p_0^{\prime})}{{\rm d} p_{0}^{\prime}} =
- \frac{3}{4 \pi} \, \frac{\omega_{pl}^2}{g^2}.
\]
Now we substitute instead of $\cos \phi^{\prime}$ the expression similar
(\ref{eq:nnn}) in (\ref{eq:rrrr}). Next integrating over $\theta^{\prime}$ we obtain
\[
\lim_{\vert {\bf k} \vert \rightarrow 0}
\frac{-i S_{k,k_1}^{(II)}}{\vert {\bf k} \vert} \bigg{\vert}_{\omega =
\omega_{\bf k}^{l}, \, \omega_1 = \omega_{{\bf k}_1}^{l}} =
3g \omega_{pl}^2 \cos \alpha \, {\bf k}_1^4
v_{{\bf k}_1}^{l} \rho_{{\bf k}_1}^{l}
\bigg{\{} \frac{\omega_{pl}}{3 \vert {\bf k}_1 \vert} -
\frac{1 - v_{{\bf k}_1}^{l} \rho_{{\bf k}_1}^{l}}
{v_{{\bf k}_1}^{l} \rho_{{\bf k}_1}^{l}}
\big[ \rho_{{\bf k}_1}^{l}( 1 - ( \rho_{{\bf k}_1}^{l})^2)
\]
\begin{equation}
(1 - F(- \rho_{{\bf k}_1}^{l})) +
\frac{{\bf k}_1^2}{3 \omega_{pl}^2} v_{{\bf k}_1}^{l}
(1 - (v_{{\bf k}_1}^{l})^2) \big] \bigg{\}}.
\label{eq:tttt}
\end{equation}
Substituting (\ref{eq:eeee}) and (\ref{eq:tttt}) in (\ref{eq:wwww}),
we find unknown expression for
$\Lambda^{S}(0, {\bf k}_1)$.

Now we return to the expression for $\gamma^{l}(0)$ (\ref{eq:bbb}), rewriting it as
follows
\begin{equation}
\gamma^{l}(0) = \int \, Q({\bf k}_1) W_{{\bf k}_1}^{l} \, d{\bf k}_1,
\label{eq:yyyy}
\end{equation}
where
\begin{equation}
Q({\bf k}_1)=
\frac{\omega_{pl} - \omega_{{\bf k}_1}^{l}}{\omega_{{\bf k}_1}
^l} \, \int \,d^{4}p \, Q_{p}(0,{\bf k}_1)p_0 \frac{{\rm d}{\cal N}_{eq}(p_0)}
{{\rm d}p_0} =
\frac{3 \omega_{pl}^2}{4 \pi g^2} \,
\frac{\rho_{{\bf k}_{1}}^{l}}{v_{{\bf k}_1}^{l}} \,
\int_0^{2 \pi}  d \varphi \,
\int_{-1}^{1}  d ( \cos \theta) Q_{p}(0, {\bf k}_1).
\label{eq:uuuu}
\end{equation}
Let us consider the integrals over angles $\varphi$ and $\theta$
in the kernel $Q({\bf k}_1)$ (\ref{eq:uuuu}). These angles enter into $\Lambda^{\Sigma}
(0,{\bf k}_1)$  and $\delta$-function in (\ref{eq:mmm}), and the element $\Lambda^{S}
(0,{\bf k}_1)$ is independed from those at all.

Now we explicitly select the angular dependence on square of the scattering
amplitude entering in $Q_{p}(0,{\bf k}_1)$
\[
\vert \Lambda^{\Sigma}(0,{\bf k}_1) + \Lambda^{S}(0,{\bf k}_1) \vert^2 =
g^4 \cos^{2} \, \phi \, {\bf k}_1^2 (v_{{\bf k}_1}^{l} \cos \theta - 1)^2 +
\]
\[
+ 2 g^2 \cos \phi \, \vert {\bf k}_1 \vert (v_{{\bf k}_1}^{l} \cos \theta - 1)
{\rm Re} \, \Lambda^{S}(0,{\bf k}_1) +
\vert \Lambda^{S}(0,{\bf k}_1) \vert^2.
\]
Using the formula of the angles connection (\ref{eq:nnn}), and taking into account
relations
\[
\int_{0}^{2 \pi}  \cos^2 \phi  d \varphi =
\pi [(3 \cos^2 \theta - 1) \cos^2 \alpha + 1 - \cos^2 \theta ] \; ,\;
\int_0^{2 \pi}  \cos \phi  d \varphi =
2 \pi \cos \theta \cos \alpha,
\]
we integrate over $\varphi$ in (\ref{eq:uuuu}). The remaining integral over $\theta$
by the $\delta$-function is computed elementary. Summing preceding, we
find
\[
Q({\bf k}_1) = \frac{\pi N_{c} g^2}{\omega_{pl}} \,
\frac{{\bf k}_1^2 \rho_{{\bf k}_1}^{l}}{\omega_{{\bf k}_1}^{l}} \,
\bigg( \frac{1 - v_{{\bf k}_1}^{l} \rho_{{\bf k}_1}^{l}}
{( \omega_{{\bf k}_1}^{l})^2 - {\bf k}_1^2} \bigg)^2
\bigg( \frac{\partial {\rm Re} \, \varepsilon^{l}(k_1)}
{\partial \omega_1} \bigg)^{-1}_{\omega_1 = \omega_{{\bf k}_1}^{l}}
\Big[ \frac{3}{4} \{ (3( \rho_{{\bf k}_1}^{l})^2 - 1) \cos^2 \alpha +
\]
\begin{equation}
+ 1 - ( \rho_{{\bf k}_1}^{l})^2 \} - 3 \rho_{{\bf k}_1}^{l} \cos^2 \alpha \,
{\rm Re} \, \tilde{\Lambda^{S}}( \vert {\bf k}_1 \vert ) +
\frac{3}{2} \cos^2 \alpha \, \vert \tilde{\Lambda^{S}}( \vert {\bf k}_1 \vert )
\vert^2 \Big] \theta (1 - \rho_{{\bf k}_1}^{l}).
\label{eq:iiii}
\end{equation}
Here, instead of $\Lambda^{S} (0, {\bf k}_1)$ we introduce a new function
depending only on $\vert {\bf k}_1 \vert$ by means of relation
\[
\Lambda^{S}(0,{\bf k}_1) =
g^2 \vert {\bf k}_1 \vert \cos \alpha \, (1 - v_{{\bf k}_1}^{l}
\rho_{{\bf k}_1}^{l}) \tilde{\Lambda}^{S}( \vert {\bf k}_1 \vert ).
\]

The pecularity of obtained expression (\ref{eq:iiii}) is the absence of the angle
dependence on $\beta$. This enables us to represent $\gamma^l(0)$ in the form
\begin{equation}
\gamma^l(0) = - 2 \pi \int_0^{\infty}  \vert {\bf k}_1 \vert^2
d \vert {\bf k}_1 \vert \int_{-1}^{1} \, d ( \cos \alpha ) \, Q ({\bf k}_1)
W_{{\bf k}_1}^l.
\label{eq:oooo}
\end{equation}

If we consider that the isotropy of the oscillations directions takes place
at the time interval which is much less than the characteristic that of the
nonlinear interaction, then the spectral density $W_{{\bf k}_1}^l$ can be
considered isotropic with respect to ${\bf k}_1$ directions. This enables
us to integrate over angle in (\ref{eq:oooo}) to completion. Now we
introduce the spectral function
\[
W_{\vert {\bf k}_1 \vert}^l \equiv
4 \pi \vert {\bf k}_1 \vert^2 W_{{\bf k}_1}^l,
\]
such that the integral $\int_0^{\infty} \, W_{\vert {\bf k}_1 \vert}^l d
\vert {\bf k}_1 \vert = W^l$ is total energy of longitudinal
oscillations in QGP. Substituting (\ref{eq:iiii}) into (\ref{eq:oooo}) and integrating over
$d( \cos \alpha)$ we find more suitable form for estimate $\gamma^l(0)$
\begin{equation}
\gamma^l(0) = - \, \int_0^{\infty} Q( \vert {\bf k}_1 \vert )
W_{\vert {\bf k}_1 \vert}^l d \vert {\bf k}_1 \vert,
\label{eq:pppp}
\end{equation}
where
\begin{equation}
Q( \vert {\bf k}_1 \vert) =
\frac{\pi N_c g^2}{2 \omega_{pl}} \,
\frac{{{\bf k}_1}^2 \rho_{{\bf k}_1}^l (1 - v_{{\bf k}_1}^l \rho_{{\bf k}_1}^l)
^2}
{[( \omega_{{\bf k}_1}^l)^2 - {\bf k}_1^2][3 \omega_{pl}^2 -
(( \omega_{{\bf k}_1}^l)^2 - {\bf k}_1^2)]}
\Big{\{} 1 - 2 \rho_{{\bf k}_1}^l {\rm Re} \tilde{\Lambda}^{S}( \vert {\bf k}_1
\vert ) + \vert \tilde{\Lambda}^S( \vert {\bf k}_1 \vert ) \vert^2 \Big{\}},
\label{eq:aaaa}
\end{equation}
\[
\tilde{\Lambda}^S( \vert {\bf k}_1 \vert )=
\frac{1}{{\bf k}_1^2 + 3 \omega_{pl}^2[1 - F(- \rho_{{\bf k}_1}^l)]}
\bigg[3 \omega_{pl}^2 \rho_{{\bf k}_1}^l
\Big( 1 - F(- \rho_{{\bf k}_1}^l) + \frac{v_{{\bf k}_1}^l}
{3(1 - ( \rho_{{\bf k}_1}^l)^2)}
\Big( \frac{{\bf k}_1^2}{\omega_{pl}^2} \,
\frac{1 - (v_{{\bf k}_1}^l)^2}{\rho_{{\bf k}_1}^l} -
\]
\[
- \frac{\omega_{pl}}{\vert {\bf k}_1 \vert} \,
\frac{1}{1 - v_{{\bf k}_1}^l \rho_{{\bf k}_1}^l} \Big) \Big) +
\frac{2 \omega_{pl}}{1 - ( \rho_{{\bf k}_1}^l)^2} \,
\Big( \vert {\bf k}_1 \vert - \frac{\omega_{pl}^2}
{\vert {\bf k}_1 \vert (1 - v_{{\bf k}_1}^l \rho_{{\bf k}_1}^l)} \Big) \bigg].
\]
Here, we set $\theta$-function entering in (\ref{eq:iiii}) equals to unit, since the
function $\rho_{{\bf k}_1}^l$ by its definition, for any $\vert {\bf k}_1 \vert
$ satisfies the inequality
\[
\rho_{{\bf k}_1}^l \leq 1.
\]

For estimate of $\gamma^l(0)$ order in (\ref{eq:pppp}) we cutt-off upper
integration limit on the characteristic value $\vert {\bf k}_1 \vert \approx
\omega_{pl}$. As was shown in [13] in the region $\vert {\bf k}_1 \vert
\geq \omega_{pl}$ there is the strong damping of longitudinal oscillations,
therefore here, this cutting has meaning. Besides, the realized above
analysis of separation of leading in $g$ terms is obeyed only in the
 long-wavelenght spectrum region. Let us approximate the kernel $Q(
\vert {\bf k}_1 \vert)$, using the approximations
\[
\omega_{{\bf k}_1}^l \approx \omega_{pl} \; , \;
\rho_{{\bf k}_1}^l \approx \frac{3 \vert {\bf k}_1 \vert}{10 \omega_{pl}}
\; , \; v_{{\bf k}_1}^l \approx \frac{\omega_{pl}}{\vert {\bf k}_1 \vert}
\; \; etc ..
\]
Leaving the leading in $ \vert {\bf k}_1 \vert $ term in the expansion (\ref{eq:aaaa}),
we obtain
\begin{equation}
Q( \vert {\bf k}_1 \vert) \approx \frac{3 \pi}{40} N_c g^2 \frac{
\vert {\bf k}_1 \vert}{( \omega_{pl})^4}.
\label{eq:ssss}
\end{equation}
Note that here, the contribution to $Q( \vert {\bf k}_1 \vert)$ from the Compton
scattering process proves to be negligible by comparison with two
other contributions. The function $W_{{\bf k}_1}^l$ is
approximated by its equilibrium value: $W_{{\bf k}_1}^l \approx
4 \pi T$ and therefore
\begin{equation}
W_{\vert {\bf k}_1 \vert}^l \approx 16 \pi^2 \vert {\bf k}_1 \vert^2 T.
\label{eq:dddd}
\end{equation}
Substituting (\ref{eq:ssss}) and (\ref{eq:dddd}) in (\ref{eq:pppp}), we finally define
\begin{equation}
\gamma^l(0) \approx - 9N_c g^2 T.
\label{eq:dddf}
\end{equation}

{\bf 10. THE GAUGE DEPENDENCE}\\

In this section we consider the problem on the gauge dependence of nonlinear
Landau damping rate of longitudinal oscillations. Let us compare derived
expression for $\gamma^{l}({\bf k})$ (\ref{eq:jjj}) with the kernel (\ref{eq:kkk}) in
covariant gauge, with similar expression computed
in temporal gauge $A_{0}^{a}=0$.

As we have mentioned in Introduction, the first nonlinear Landau damping rate
in temporal gauge was calculated in [8]. The inaccuracies in calculations
were made in obtaining of equation for the second correction of a gauge
field $A^{(2)}(k)$ (formula (3.19) in [8]). The elimination of these inaccuracies
leads to the expression
\[
( \omega^{(0)})^2 \epsilon( \omega^{(0)},{\bf k}) A^{(2)}(k) =
g \sum_{k_1 + k_2 = k} \frac{{\bf k} \cdot {\bf k}_1}{K_1}
\frac{{\bf k} \cdot {\bf k}_2}{K_2} \frac{1}{K}
[A^{(1)}(k_1),A^{(1)}(k_2)] -
\]
\begin{equation}
- g^3 \int \frac{d^3 p}{(2 \pi)^3 E_{p}}
\Big[ N_{f}( \frac{df^{(0)}(p)}{dE_{p}} +
\frac{d{\bar f}^{(0)}(p)}{dE_{p}}) +
2N_{c} \frac{dG^{(0)}(p)}{dE_{p}} \Big] \frac{1}{p \cdot k^{(0)} +
ip^{0}0^{+}}
\label{eq:ffff}
\end{equation}
\[
\sum_{k_1 + k_2 = k} \Big \{ \frac{1}{2p^{0}} \, \frac{{\bf p} \cdot
{\bf k}_2}{p \cdot k_{2}^{(0)} + ip^{0}0^{+}} \Big \}
\frac{{\bf p} \cdot {\bf k}}{K}
\frac{{\bf p} \cdot {\bf k}_1}{K_1}
\frac{{\bf p} \cdot {\bf k}_2}{K_2}
[A^{(1)}(k_1),A^{(1)}(k_2)].
\]
Here, we use notations accepted in [8]. The distinction (\ref{eq:ffff}) from (3.19)
in [8] is as follows. Firstly - this is the availability of the first  term in
the right-hand side of (\ref{eq:ffff}), that is connected with self-action of a gauge
field, and is of the same order in a soft region of excitations as the second
term, connected with medium effects. As shown above this term also
contributes to process of the nonlinear scattering of plasmons by QGP particles.
Second distinction lies in fact that instead of expression in braces
in the second term in the right-hand side of Eq. (\ref{eq:ffff}) in [8] the following
expression
\[
\frac{\omega_2^{(0)}}{p \cdot k_2^{(0)} + ip^{0} 0^{+}}
\]
is used.
The principal point is presence of the factor 1/2 in our case. This factor
enables us finally to lead the expression for nonlinear Landau damping rate
in the temporal gauge to
the form that is similar to (\ref{eq:aaaa}), (\ref{eq:ssss}). Further we obtain the expression
for $\gamma^{l}({\bf k})$ in this gauge following our reasoning, do not repeating
calculations in [8].

In temporal gauge the Yang-Mills equation (\ref{eq:w}) has the form
\[
\partial_{\mu}F^{\mu \nu}(x) - ig[A_{\mu}(x),F^{\mu \nu}(x)] -
\xi^{-1}u^{\nu}u^{\mu}A_{\mu}(x) = -j^{\nu}(x).
\]
Here, as the fixed four-vector in the gauge condition $n_{\mu}A^{\mu}=0$
we choose the four-velocity of plasma: $n_{\mu} \equiv u_{\mu}$.

In this gauge instead of the propagator (\ref{eq:vv}) we have
\[
{\cal D}_{\mu \nu}(k) =
-[k^2g_{\mu \nu} - k_{\mu}k_{\nu} + \xi^{-1}u_{\mu}u_{\nu} - \Pi_{\mu \nu}(k)
]^{-1} =
\]
\begin{equation}
= - \, \frac{P_{\mu \nu}}{k^2 - \Pi^{t}(k)} -
\frac{\tilde{Q}_{\mu \nu}}{k^2 - \Pi^{l}(k)} -
\xi \frac{k^2}{(ku)^2}D_{\mu \nu}.
\label{eq:gggg}
\end{equation}
In obtaining of the last equility in (\ref{eq:gggg}) we use relation [15]
\[
u_{\mu}u_{\nu} = \frac{\bar{u}^2}{k^4}Q_{\mu \nu} -
\frac{(ku)}{k^4} \sqrt{-2k^2 \bar{u}^2}C_{\mu \nu} +
\frac{(ku)^2}{k^2}D_{\mu \nu}
\]
and introduce notation
\[
\tilde{Q}_{\mu \nu} \equiv Q_{\mu \nu} +
\frac{\sqrt{-2k^2 \bar{u}^2}}{k^2(ku)} C_{\mu \nu} +
\frac{\bar{u}^2}{k^2(ku)^2}D_{\mu \nu},
\]
where $C_{\mu \nu} = -( \bar{u}_{\mu}k_{\nu} + \bar{u}_{\nu}k_{\mu})/
\sqrt{-2k^2 \bar{u}^2}.$ In the rest frame of a plasma, tensor
$\tilde{Q}_{\mu \nu}$ has the structure
\[
\tilde{Q}_{\mu \nu} =  - \frac{k^2}{\omega^2} \left(
\begin{array}{cc}
0 & 0 \\
0 & \frac{{\bf k} \otimes {\bf k}}{{\bf k}^2}
\end{array}
\right) .
\]
Further we take, $\xi = 0$, i.e. $A_0^{a}=0$
is imposed strictly. By virtue of expansion of the polarization tensor
$\Pi_{\mu \nu} = P_{\mu \nu} \Pi^{t} + Q_{\mu \nu} \Pi^{l}$ and the properties
of tensor structures $P,Q,C$ and $D$ [15], the expression $\Pi^{l}$ remains
unchanged, i.e.
\[
\Pi^{l} = Q^{\mu \nu} \Pi_{\mu \nu} \equiv \tilde{Q}^{\mu \nu} \Pi_{\mu \nu}.
\]
Instead of expansion of the spectral density (\ref{eq:cc}) now we have
\[
I_{\mu \nu} = I_{k}^{t} \, P_{\mu \nu} + I_{k}^{l} \, \tilde{Q}_{\mu \nu}.
\]
Connection between the spectral density $I_{\bf k}^{l}$ and the density of the energy
of longitudinal oscillations is unchanged.

From the above it is easy to appreciate that for obtaining of nonlinear
Landau damping rate in the temporal gauge it will sufficiently in the expression (7.9)
to replace the projector $Q$ by $\tilde{Q}$, and in the expression for $\tilde
{\Sigma}$ (\ref{eq:zz}) use propagator (\ref{eq:gggg}) instead of (\ref{eq:vv}). Finally we obtain
the same expression (\ref{eq:jjj}), where now in the kernel (\ref{eq:kkk}) instead of the
scattering amplitudes $\Lambda^{\Sigma}({\bf k},{\bf k}_1),
 \Lambda^{S^{(I)}}({\bf k},{\bf k}_1)$ and
$\Lambda^{S^{(II)}}({\bf k},{\bf k}_1)$ it is necessary to use expressions
\[
\Lambda_{0}^{\Sigma}({\bf k},{\bf k}_1) =
\frac{g^2}{\vert {\bf k} \vert \vert {\bf k}_1 \vert} \,
\frac{({\bf k}{\bf v})({\bf k}_1 {\bf v})}{\omega_{\bf k}^{l} -
({\bf k}{\bf v})} \,
\frac{[( \omega_{\bf k}^{l})^2 - {\bf k}^2]
[( \omega_{{\bf k}_1}^{l})^2 - {\bf k}_1^2]}
{\omega_{\bf k}^{l} \, \omega_{{\bf k}_1}^{l}},
\]
\[
\Lambda_{0}^{S^{(I,II)}}({\bf k}, {\bf k}_1) =
\frac{g}{\vert {\bf k} \vert \vert {\bf k}_1 \vert {\bf k}_2^2}
\left( \frac{1}{k_2^2} \, \frac{-iS_{0 \, k,k_1}^{(I,II)}}
{\varepsilon^{l}(k_2)} \right)_{\omega = \omega_{\bf k}^{l}, \,
\omega_1 = \omega_{{\bf k}_1}^{l}},
\]
respectively.
Here,
\[
-iS_{0 \, k,k_1}^{(I)} = 0,
\]
\[
-iS_{0 \, k, k_1}^{(II)} = -g^3 \frac{k^2 k_1^2 k_2^2}{\omega \omega_1 \omega_2}
\int d^4 p^{\prime} \, \frac{
({\bf p}^{\prime}{\bf k})
({\bf p}^{\prime}{\bf k}_1)
({\bf p}^{\prime}{\bf k}_2)}{p^{\prime}k_2 + i p^{\prime}_{0} \epsilon}
\left(
\frac{(k \partial_{p^{\prime}}{\cal N}_{eq})}{p^{\prime}k} -
\frac{(k_1 \partial_{p^{\prime}}{\cal N}_{eq})}{p^{\prime}k_1} \right).
\]

Now we compare the scattering amplitude $\Lambda$ in the covariant gauge with
appropriate scattering amplitude $\Lambda_{0}$ in $A_{0}$-gauge. Using
the above-mentioned explicit expressions for amplitudes, as the result of
simple calculations we obtain
\begin{equation}
\Lambda - \Lambda_0 = \frac{g^2}{\vert {\bf k} \vert
\vert {\bf k}_1 \vert}
\frac{k^2 k^2_1 - \omega^2 \omega^2_1}{2 \omega \omega_1}
({\bf k} + {\bf k}_1)({\bf v} - {\bf v}_{\parallel}) \neq 0,
\label{eq:hhhh}
\end{equation}
where ${\bf v}_{\parallel} \equiv {\bf k}_2 ({\bf k}_2 {\bf v})/ {\bf k}_2^2$.
Here the term with ${\bf v}_{\parallel}$ in the right-hand side depends on
the contribution to the scattering amplitudes the terms with longitudinal
virtual oscillation. The last expression explicitly demonstrates
the gauge-noninvariant character of obtained decrement of nonlinear Landau
damping.\\

{\bf 11. CONCLUSION}\\

In our paper it was shown that the nonlinear interaction of longitudinal
eigenwaves leads to effective pumping of energy across the spectrum sideways
of small wave numbers.
Consequence of this fact is the inequality: $\gamma^l(0) < 0$, i.e.
 ${\bf k}=0$ -
mode is increased. The real nonlinear absorption of the energy of plasma waves
by particles in QGP is the effect of higher-order.

It is clear that growth of ${\bf k}=0$ - mode is consequence of chosen
approximation. In the region of a small $\vert {\bf k} \vert$ effects,
described by nonlinear terms in the expansion of the color current of
higher-order in the field, come into play .
One of such possible nonlinear effects is known
from the theory of electromagnetic plasma [11]. This is as follows.

By the effect of pumping, all plasmons will be tend to concentrate near a
small $\vert {\bf k} \vert = \vert {\bf k}_0 \vert \rightarrow 0$. However,
phase space, which the plasmons are occupied, proportionaled to
$\vert {\bf k}_0 \vert^3$ will be also highly small. By virtue of this fact
the intensive collision of plasmons is arised. This is the process
\[
l + l_1  \rightleftharpoons l^{\prime} + l_1^{\prime},
\]
which is to lead to the scattering of plasmons from region of a small
$\vert {\bf k} \vert$ and thus to suppression of increase of ${\bf k}=0$-mode.
The probability of this four-plasmon interaction is defined by the preceding
method from the nonlinear current of fourth order $j_{\mu}^{T(4)}$ with
regard to the process of interaction iteration of higher-order in field.

Let us consider in more detail approximations scheme, which we use in this
paper. In fact, here two various levels of the approximations are used. The first
of them is connected with the employment of usual approach, developed
in EMP to QGP, i.e. the standard approximation of the current in terms of
the oscillations amplitude and computation of interacting field in the form of a
series of a perturbation theory in a free field. However, in contrast to
EMP, in our case even if the first two nonlinear orders of the color current
are taken into account, vastly more terms, defining the nonlinear scattering of
waves are derived. Here we use second approximation level,
connected with the notions going from the papers on hard thermal loops or
more precisely, the set of the orders estimates of a various terms, developed
by Blaizot and Iancu [5]. This set of estimates enables us to extract
the leading terms in the coupling constant from the set of obtained
terms. The surprising thing is that although this terms are purely non-Abelian,
all basic conclusions, performed on the basis of these terms, coincide with
the appropriate results in nonlinear theory of EMP in a qualitative sense.

The obtained expression of nonlinear Landau damping rate (\ref{eq:jjj}),
(\ref{eq:kkk})
is not gauge invariant, that it was
explicitly shown in Sec.10.
Such gauge dependence of the nonlinear damping rate is a specific character of
the non-Abelian theory. In the Abelian plasma a similar problem does not arise,
since the theory of the nonlinear processes in EMP is stated in terms of the
gauge-invariant electric and magnetic fields only. However in the case of QGP
it is impossible to avoid the work with potentials, since ones explicitly
appear in the kinetic equations (\ref{eq:r}) and YM Eq. (\ref{eq:w}). Therefore the initial
function in our consideration is the two-point correlative function of the form
(\ref{eq:v}) (more precisely, its longitudinal part, since in this paper we have restricted
ourselves to the longitudinal excitations in QGP only), weakly inhomogeneous
and weakly nonstationary, having no directly physical meaning and being the
gauge-noncovariant value.

By virtue of fact that the process of the nonlinear scattering of
longitudinal waves is the physical process, the scattering amplitude
$\Lambda$, entering in the kernel (\ref{eq:kkk}) is bound to be gauge-invariant
(within used in this paper approximation the spectral density of energy
$W^{l}_{{\bf k}_1}$ is gauge-invariant value).
However, as it was shown in Sec.10 (Eq.(\ref{eq:hhhh})), the scattering amplitudes
calculated in the covariant and $A_0$-gauges are not coincident.
The simplest analysis of the right-hand side of (\ref{eq:hhhh}) points to the fact
that difference $\Lambda - \Lambda_0$ is equal to zero if in the last
parentheses in the right-hand side besides longitudinal in relation to the
vector ${\bf k}_2$ component of velocity ${\bf v}_{\parallel}$ the transverse
component ${\bf v}_{\perp}$: $({\bf v}_{\perp} {\bf k}_2) = 0$ is present.
Intuitively clear that such term is arised from discarded contribution to
nonlinear scattering of term with transverse virtual oscillation.
Howerever if the longitudinal and transverse virtual oscillations taken
into account, already it is impossible to write $\gamma^{l}
({\bf k})$ in compact and transparent form identical with those of (\ref{eq:jjj}),
(\ref{eq:kkk}). Therefore in this case checking of gauge invariance becomes a very
nontrivial problem and it is the subject of specific research.

\section*{\bf Acknowledgements}

This work was supported by the Russian Foundation for Basic Research
(Project No.97-02-16065).

\end{document}